INTERNSHIP REPORT

# Design and Construction of the Test-Stand for Split and Delay Line at European XFEL

17th August 2015

**Hong XU**[1]

Supervisor:
**Wei Lu**[2,3], **Anders Madsen**[2], **Sergio Di Matteo**[1]

[1]Université de Rennes 1, Rennes, France
[2]European XFEL Facility GmbH, Hamburg, Germany
[3]Technische Universität Berlin, Berlin, Germany

Materials Imaging and Dynamics Group (MID)

European X-ray Free Electron Laser Facility GmbH

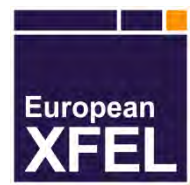

# Contents

# 1 Introduction

This chapter is the general introductions to the European X-ray Free Electron Laser (European XFEL), the Deutsches Elektronen-Synchrotron (DESY), and also the internship report contents.

## 1.1 European X-ray Free Electron Laser

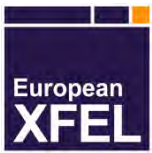

The European XFEL is one of the world's most advanced large scale research facilities currently under construction in Hamburg, Germany. It will generate extremely intense and ultrashort X-ray flashes that will be used by researchers from all over the world from 2017 on[1].

Since 12 countries already participated in the European XFEL project, a limited liability non-profit company was established in October 2009, which is called European X-ray Free Electron Laser Facility GmbH with DESY as the largest shareholder. The construction costs of this facility, which includes the commissioning, amount to 1.15 billion Euros (price levels of 2005)[2].

To generate X-ray flashes with outstanding peak brilliance, the electrons need to be brought to high energies through accelerators, and then induce them to emit light. The designed construction to realize this great ambition is shown in figure 1-1[1]. The electron bunches are generated by knocking the electrons out of a piece of metal using a conventional laser. Then through the injector in DESY campus site, it will enter the 1.7 km particle accelerators placed in the tunnels 6 to 38 m underground with inner diameters of up to 5.3 m. The superconducting resonators are cooled down to -271 degree Celsius, and the electrical current flows through the resonators without any loss and the oscillating microwave can transfer its entire energy to the electrons[1].

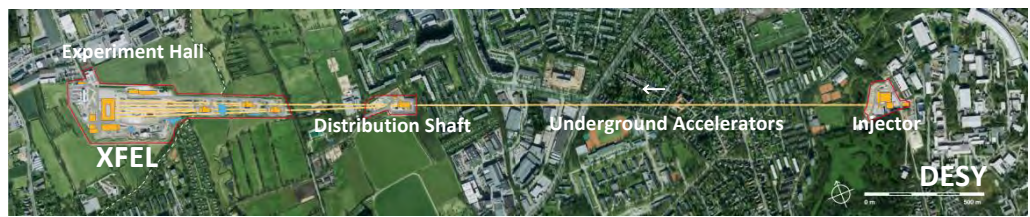

Figure 1-1: Aerial view of the European XFEL facility[1]. Total length 3.4 km.
Right to left: DESY-Bahrenfeld site with injector, Osdorfer Born site with distribution shaft, and the XFEL-Schenefeld Campus site with underground experiment hall, large laboratory and office building on the top.

After that, the accelerated electrons will pass through the so-called undulators with more than 100m length, which are magnet arrays to force the electrons onto a tight slalom course and emit X-ray radiation. Since the radiation is faster than the electrons speeding along the slalom path, the radiation will overtake the electrons beam and interact with them. This process induces the micro branching of electrons by accelerating some electrons and slowing others down. Therefore the achieved X-ray flashes are extremely short and intense and all the electrons in the same micro branching will emit light synchronously. This process is called self-amplified spontaneous emission (SASE)[2].

Depending on the wavelength and intensity requirements of different beamline, the x-ray flashes are distributed to different end user group as shown in figure 1-2[3]. Now there are six end stations and their instruments are under construction.

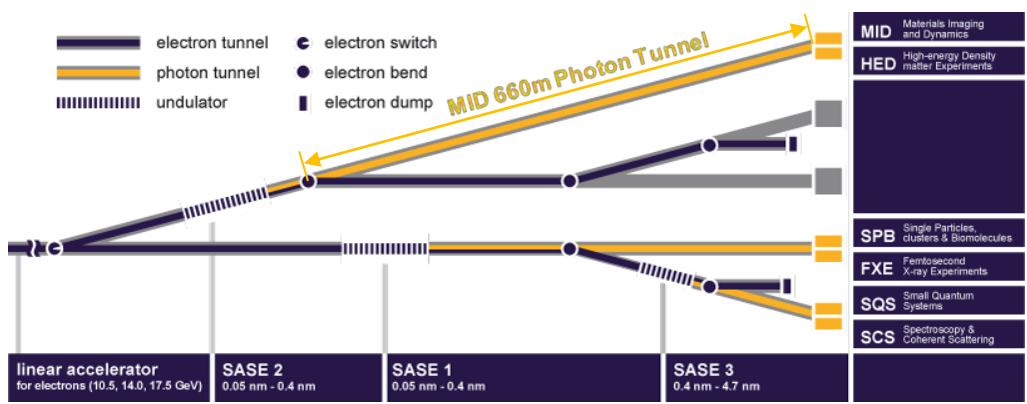

Figure 1-2: Arrangement of research beamlines in European XFEL[3].
Different SASE x-ray wavelength can be achieved by changing the magnetic fields in undulators.

**1. MID: Materials Imaging and Dynamics.**

The aim of the MID group is to enable studies of structure and dynamics in condensed matter by means of X-ray scattering and imaging experiments on nanoscale[4]. Mechanical and structural designs of MID beamline are in process. This is the group where I stayed for three months internship and my main work includes design and construction of the test-stand for split delay line.

**2. HED: High Energy Density Matter Experiments.**

The HED instrument will be a unique platform for experiments combining hard X-ray free-electron laser radiation and the capability to place matter under extreme conditions of pressure, temperature, or electric field using high-energy optical lasers or pulsed magnets. Scientific applications will include studies of the properties of matter within exoplanets, new extreme-pressure phases and solid-density plasmas, and structural phase transitions of complex solids in high magnetic fields[2].

**3. SPB: Single Particles Clusters & Biomolecules.**

The SPB group aims to deliver a state-of-the-art single-particle imaging instrument for the European XFEL. Imaging capabilities will cover the whole range of single particles—from larger biological particles, such as cells and organelles, to certain material particles and nanocrystalline biomolecules, with aspirations to also image individual, biologically relevant molecules[2].

**4. FXE: Femtosecond X-ray Experiments.**

The FXE instrument will enable time-resolved pump–probe experiments on ultrafast time scales for a broad scientific community. While the first components have already been manufactured, the FXE group is furthering the design of the remaining components in research campaigns using laboratory laser sources, synchrotron sources combined with the FXE MHz laser system, and free-electron lasers (FELs)[2].

**5. SQS: Small Quantum Systems.**

The SQS group focused on defining the experiment area layout, including the final design of the focusing optics, the hutch infrastructure, and the integration of specific user instrumentation. The group also continued its science programme dedicated to the application of two-colour experiments for studying atomic and molecular photoionization dynamics and to investigations of non-linear processes in prototype atomic samples with experiments at other free-electron laser sources[2].

**6. SCS: Spectroscopy & Coherent Scattering.**

The SCS instrument will be dedicated to the study of electronic, spin, and atomic structures on the nanoscale using soft X-rays. Its purpose is to enable users to explore excited-state dynamics on ultrafast time scales and to unravel the function of complex materials. Areas of application are materials science, nanoscience, and condensed-matter dynamics as well as bioscience[2].

**The Civil Construction:**

In 2014, civil construction began to enter its final phase with the beginning of construction of the headquarters building in Schenefeld. By the end of last year, the first floors of the laboratories and offices, which is a three-floor structure, were already finished successfully. And also the above-ground halls on the DESY site and the Osdorfer Born sites were built. Planning for landscaping on the three sites has begun as well from last year[2]. In June 2015, I was lucky to have an opportunity of visiting XFEL Schenefeld site, during which I got the overview of the constructions there. The main body of the office building was already completed and workers were doing the internal and external decorations. Also I visited the underground photon tunnel and the MID experimental hall. The expected time for MID group to move there is July 2016.

## 1.2    Deutsches Elektronen-Synchrotron

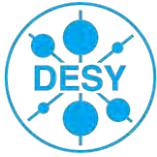

The Deutsches Elektronen-Synchrotron (DESY) is one of the world's leading research centers for the investigation of the structure of matter, which is supported by public funds and members of the Helmholtz Association. It develops, runs and uses accelerators and detectors for photon science[5] and particle physics[6].

The research of photon science at DESY mainly focus on the research of atomic and molecular dynamics; surfaces, thin films and interfaces; electronic and magnetic structure; structure and structural dynamics; macromolecular crystallography; science of X-ray sources. While for particles physics, the research topics are challenging, including HERA, LHC, linear accelerators, experimental activities, astroparticle physics and particle physics theory[6].

In the framework of DESY campus, as unique research tools, there are 17 facilities of accelerator building and running for various international projects as shown in figure 1-3[7], European XFEL included. My lab work was mainly done in HERA Süd (south of HERA ring), which was used for high energy physical experiments while now a part of it becomes underground laboratory for testing of European XFEL's projects.

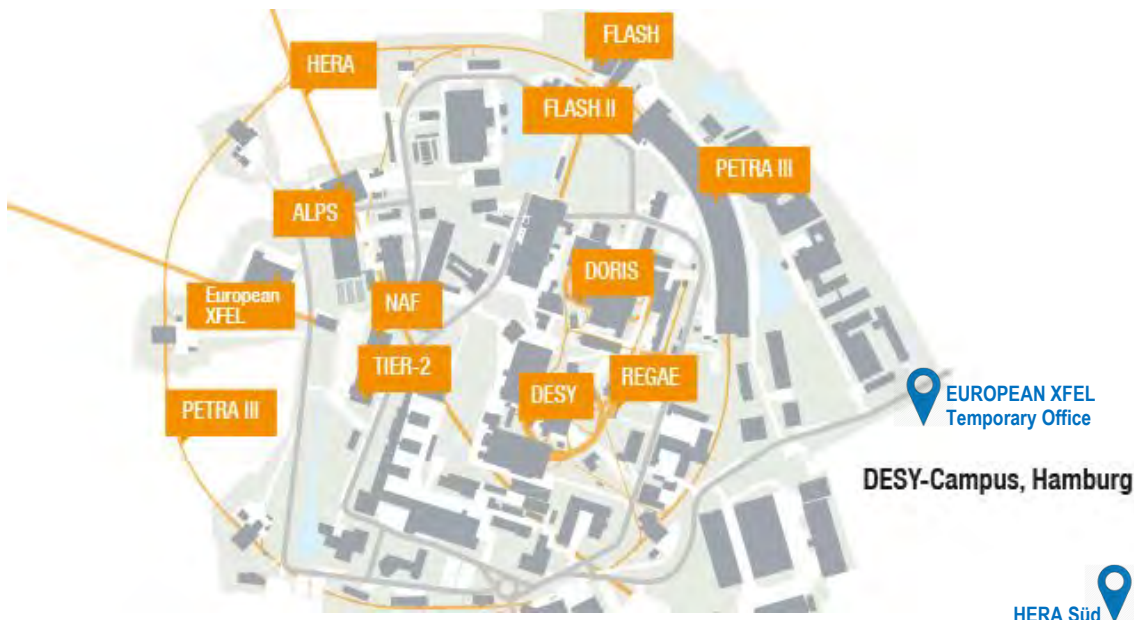

Figure 1-3: DESY Campus in Hamburg including various accelerator facilities[7].

PETRA III: The world's best storage ring generating X-ray.
FLASH: The free-electron laser generates ultra-short pulsed X-ray laser light.
REGAE: The source for relativistic electron beams enables innovative experiments.
TIER-2: The computing centre provides large computing and storage systems.
NAF: The National Analysis Facility at DESY is a powerful computing complex for particle physicists.
HERA:  The measuring data of the largest particle accelerator at DESY are still being evaluated.
DORIS: The storage ring at DESY was used for particle physics and research with synchrotron radiation.
ALPS: The experiment at DESY searches for extremely light particles.
European XFEL: The new X-ray laser is currently built in the metropolitan region of Hamburg.

## 1.3 Contents of the Internship Report

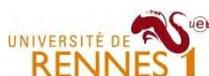

This mandatory internship report required by Université de Rennes 1 mainly presents the work I have done at European XFEL in MID group. It consists of five chapters. After the general introduction in Ch.1, Ch.2 and Ch.3 present the fundamental theory and mechanical design of split and delay line (SDL). In Ch.4 mechanical design of the test-stand for this SDL is shown. The last Ch.5 is the practical work, which is the preparation and test of the test-stand. The following are the detail descriptions of each chapter and the knowledge I learned.

In Ch.2 and Ch.3, introduction to principles, application and mechanical design of split and delay line at MID station is presented. To have a better understanding, I studied the MID and SDL Technical Design Report (TDR) by deducing formulas and reading references. Also I found the link between my knowledge and XFEL SDL, which is the crystallography I've learned in Rennes. Thus I tried to organize the main concepts that I understood from the formal TDRs, and then rewrote it in my own words in these chapters.

Ch.4 is about the mechanical design of the test-stand for SDL, which is mainly the design work I have worked with. The test-stand is aimed to partly simulate the final SDL and test its functions in high vacuum situation. I'm involved in this procedure starting from the selection and purchase of components. Since my background is materials chemistry and partly theoretical physics, sincerely I feel quite difficult with mechanical design affairs in the very beginning. But I did learn a lot of very practical and useful skills during this period, like Solid Edge software[8], electronic and optic configuration, mechanical limitation consideration and etc.

As mentioned above, the last chapter presents the preparation and test of the test-stand. I did these works with the great help from nice XFEL scientists and engineers. The preparation and test were mainly done in Hera Süd underground experimental hall, where there are mechanical tools, cleanroom and equipment. In this chapter, what I learned is very practical for engineering domain that I never involved in. Thus I was very excited to do things like welding the D-Sub connector, assembling of flanges, test of chamber vacuum and etc. Sincerely I would say, these give me very specific cognition for engineering, which I believe has significant influence on my further scientific research.

# 2 Split and Delay Line at MID

Owing to the limitation of energy domain, the studies of fast dynamics at microsecond level or even lower are difficult for scientists. The development of XFELs techniques, which were firstly coined by John Madey[9] in 1976 at Stanford University, will make it possible to study fast dynamics with time structure by using correlation spectroscopy. Nowadays, the way to study these kinds of fast non-equilibrium dynamics is ultra-fast X-ray pump – X-ray probe experiment, for which the split and delay line (SDL) is indispensable[4].

Looking around the world, previously the SDLs were built and operated with Extreme Ultraviolet Lithography (EUV) light and soft X-rays[10], while this conception is adopted and then analogically applied into hard X-ray operation range. A prototype was built for Linac Coherent Light Source (LCLS)[11] in California, USA. Besides, there is also a SDL project[12] operating at the XFEL facility SACLA in Japan, which shares similar principles with the others. In this chapter, I mainly present the mechanical design of the Split and Delay Line in MID group at European XFEL.

## 2.1 Placement of Split and Delay Line

The SDL is placed in the MID Optics Hutch which is behind the Common SASE-2 Tunnel and the MID Photon Tunnel, as shown in figure 2-1[4].

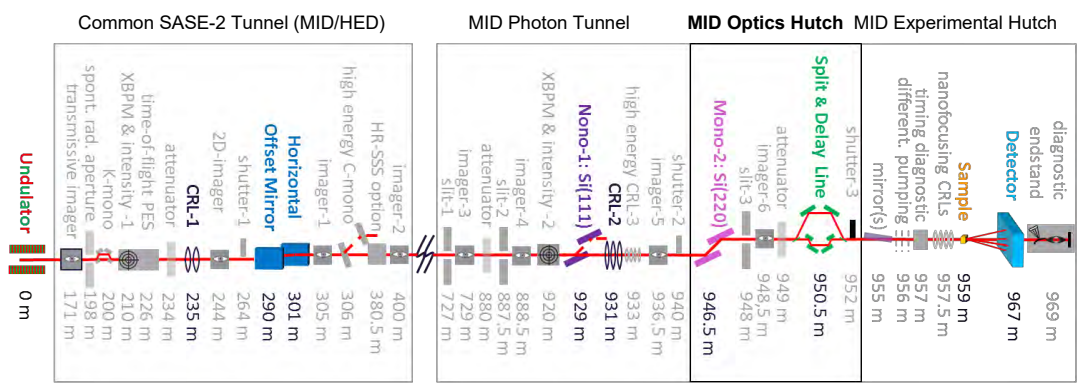

Figure 2-1: Side overview of MID beamline layout including the placement of SDL[4].

The vital considerations for this placement of SDL are the following two aspects[4]:

1. It should be placed after a monochromator to reduce the heat load on the first beam splitter crystal;
2. It should be placed as close as possible to the sample since a long lever arm is unfavorable for beam position stability.

Especially, both the pump beam and the probe beam need to overlap at the sample and the detector for many kinds of experiments. This will be explained later in the conception part of SDL.

The photons are generated in the 220m long undulator arrays with 35 undulators of 5 m magnetic length each plus space in-between. About 235m after the source point, before the offset mirrors, a beryllium refractive lens (CRLs) stack allows for moderate focusing or collimation of X-ray beam. After that, through two offset mirrors at 290m and 301m, the beam will continue its travelling until the monochromator Si (111), which aims to reduce the band width of transmitted photons. The main focusing CRL stacks at 931m and following is a medium resolution monochromator Si (220). Now it comes to the Split and Delay Line at 950.5m while the sample is placed at 959m and the detector is approximately at 967m[4].

## 2.2   Conception of Split and Delay Line

The conception of SDL is shown in figure 2-2 with two operation modes: collinear mode[a] and inclined mode[b]. The incident XFEL pulse is split into two parts by beam splitter. Half pulse travels along the upper branch while another half travels along the lower branch. The delay of half pulse is achieved by changing the height (H) of upper branch crystal. The two channel-cuts setting in the lower branch are aimed to achieve the determination of Δt=0[4]. The two pluses will merge again at the beam merger in the collinear mode, or travel on two different optical paths towards the sample in the inclined mode. For the latter option, it needs a grazing incidence mirror behind the SDL to achieve the overlapping of the pluses at the sample[4].

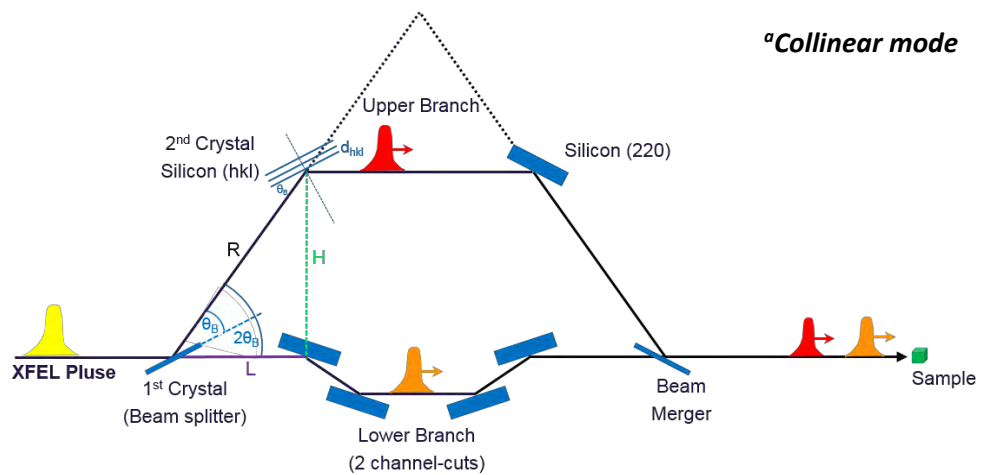

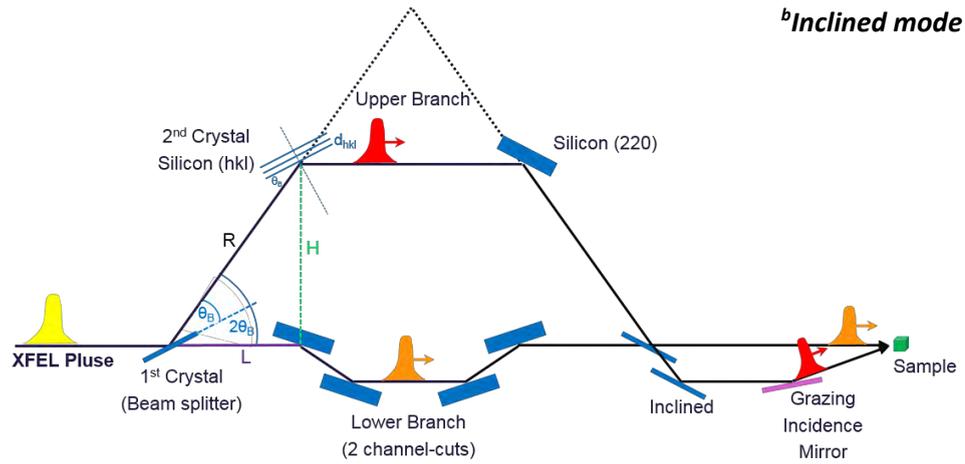

Figure 2-2: Concept of SDL with collinear mode and inclined mode
Provided by Dr. Wei Lu. H is the height of the second crystal in the upper branch.

In this way, a few femtoseconds temporal separation of X-ray pulse pairs through branches can be achieved[4]. That is significant for the ultra-fast X-ray pump – X-ray probe experiment in the investigation of ultrafast dynamics.

## 2.3   Beam Path Delay Geometry

Depending on the concept of SDL in chapter 2.2, one can deduce the beam path delay geometry. The geometry is given in above figure 2-2. In order to achieve a zero beam offset for SDL, the structure of SDL should be symmetric in the direction of beam. Thus it achieves half delay in the first half upper branch and another half delay in the second one. By this symmetric delay setting, the beam can back to the initial beam direction. Since it has the nice property of symmetric structure, only the first half of the upper branch is taken into consideration in this chapter.

Here the simple trigonometry calculation is presented:
The time delay is:
$$dt_{\frac{1}{2}} = (R - L)/c \qquad (1)$$

From the geometry sketch one can easily find that:
$$L = R\cos 2\theta_B \qquad (2)$$

Thus here:
$$dt_{\frac{1}{2}} = \frac{R-L}{c} = \frac{R(1-\cos 2\theta_B)}{c} = \frac{R(1-\cos^2\theta_B + \sin^2\theta_B)}{c} = \frac{2R\sin^2\theta_B}{c}$$

$$\Rightarrow \quad 2R\sin^2\theta_B = cdt_{\frac{1}{2}}$$

$$\Rightarrow \quad R = \frac{1}{2\sin^2\theta_B} cdt_{\frac{1}{2}}$$

$$\Rightarrow \quad L = R\cos 2\theta_B = \frac{\cos^2\theta_B - \sin^2\theta_B}{2\sin^2\theta_B} cdt_{\frac{1}{2}} = \frac{1}{2}\left(\frac{1}{\tan^2\theta_B} - 1\right) cdt_{\frac{1}{2}}$$

$$\Rightarrow \quad H = R\sin 2\theta_B = \frac{2\sin\theta_B \cos\theta_B}{2\sin^2\theta_B} cdt_{\frac{1}{2}} = \frac{1}{\tan\theta_B} cdt_{\frac{1}{2}}$$

The length of R, L and H in representation of $\theta_B$ and $dt_{1/2}$:

$$R = \frac{1}{2\sin^2\theta_B} cdt_{\frac{1}{2}} \qquad L = \frac{1}{2}\left(\frac{1}{\tan^2\theta_B} - 1\right) cdt_{\frac{1}{2}} \qquad H = \frac{1}{\tan\theta_B} cdt_{\frac{1}{2}} \qquad (3)$$

Use the expression for L with $\tan\theta_B = \frac{\sin\theta_B}{\sqrt{1-\sin^2\theta_B}}$. This gives:

$$L = \frac{1}{2}\left(\frac{1}{\tan^2\theta_B} - 1\right) cdt_{\frac{1}{2}} = \frac{1}{2}\left(\frac{1-\sin^2\theta_B}{\sin^2\theta_B} - 1\right) cdt_{\frac{1}{2}} = \left(\frac{1}{2\sin^2\theta_B} - 1\right) cdt_{\frac{1}{2}}$$

Thus:
$$\frac{cdt_{\frac{1}{2}}}{2\sin^2\theta_B} - L = cdt_{\frac{1}{2}} \qquad (4)$$

The same expression $\tan\theta_B = \frac{\sin\theta_B}{\sqrt{1-\sin^2\theta_B}}$ for H gives:

$$\sin^2\theta_B = \frac{\frac{c^2 dt_{\frac{1}{2}}^2}{H^2}}{1 + \frac{c^2 dt_{\frac{1}{2}}^2}{H^2}} \qquad (5)$$

Combine equation (4) and (5), solve H as a function of L:

$$\frac{1}{2} cdt_{\frac{1}{2}} * \frac{1 + \frac{c^2 dt_{\frac{1}{2}}^2}{H^2}}{\frac{c^2 dt_{\frac{1}{2}}^2}{H^2}} - L = cdt_{\frac{1}{2}}$$

Simplify the above equation and it gives:

$$H(L) = \sqrt{cdt_{\frac{1}{2}}(2L + cdt_{\frac{1}{2}})} \qquad (6)$$

In the SDL, Si (220) crystal will be used[4]. Figure 2-3 shows the beam path in the full upper branch as a function of beam direction for the selected Si (220) refelction[4] with photon energy of 5keV and 10keV at $dt = 2dt_{1/2} = 800$ ps[4].

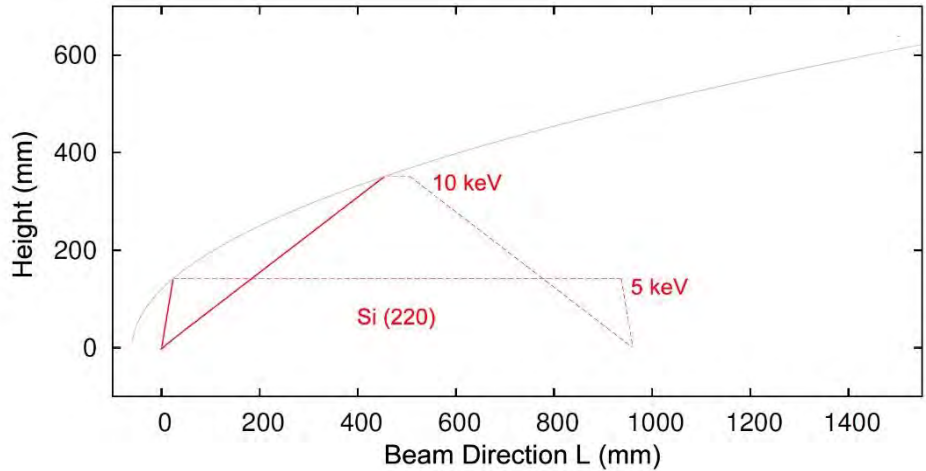

Figure 2-3: The 800 ps trajectories of the upper branch crystal for the Si (220) reflection with photon energy of 5keV and 10 keV[4].

## 2.4 Stability Consideration

### 2.4.1 Mechanics Stability

A desired focus spot at the sample or at the detector position is in a scale of 10×10 µm$^2$, which is possible by using intermediate focusing scheme with CRL stacks. Where the size of the beam on the sample under correct CRL focusing is[4]:

$$S_{FWHM}^{foc.} = S_{FWHM}^{Source} * \frac{513}{213} * \frac{28}{209} \approx [30\ to\ 60\ \mu m] * \frac{1}{3.1} \approx 10\ to\ 20\ \mu m$$

According to the placement of SDL, the sample chamber is supposed to be placed about 7m after the SDL while the detector approximately another 8m downstream. In the energy range from 5 to 10 keV, the smallest beam size that can be achieved with a stability of 1 to 2 µm at a distance of 15m, which is 10% to 20% size of the beam FWHM in focus. That means the angular stability is 0.1 µrad, which is also the requirement for motorized pitch or Bragg rotation stages[4].

### 2.4.2 Temperature Stability

The stabilization of temperature for SDL is required for special attention. In this design, the SDL will have a proper enclosure in the optics hutch that must be stabilized in temperature with a preferable level better than 0.1K. In addition, a system for temperature stabilization of the crystal cages is necessary[4].

### 2.4.3 Vacuum Requirement

The vacuum requirement for the tunnels' beam transfer system and MID optics hutch is around 10$^{-7}$ to 10$^{-8}$ mbar and the same requirement for the SDL. Moreover, the optical bench will be built in a large vacuum vessel, which should provide easy intervention and be vibration free[4].

# 3    Mechanical Design of SDL

This chapter presents the detail and latest mechanical designs of split and delay line at MID. Most of these works were previously done by Dr. Tino Noll from TU Berlin and Dr. Wei Lu.

## 3.1    Upper and Lower Branch

Mechanical designs of SDL are based on the formal explained principle and the mechanical structure of upper and low branch is shown in figure 3-1. There are several designs of upper and lower branch, while the final decision is the L shape optical bench which is possible to bend or weld plates together followed by annealing. And it should be mounted with free strain onto the support structure using a three pillar mount and bellows.

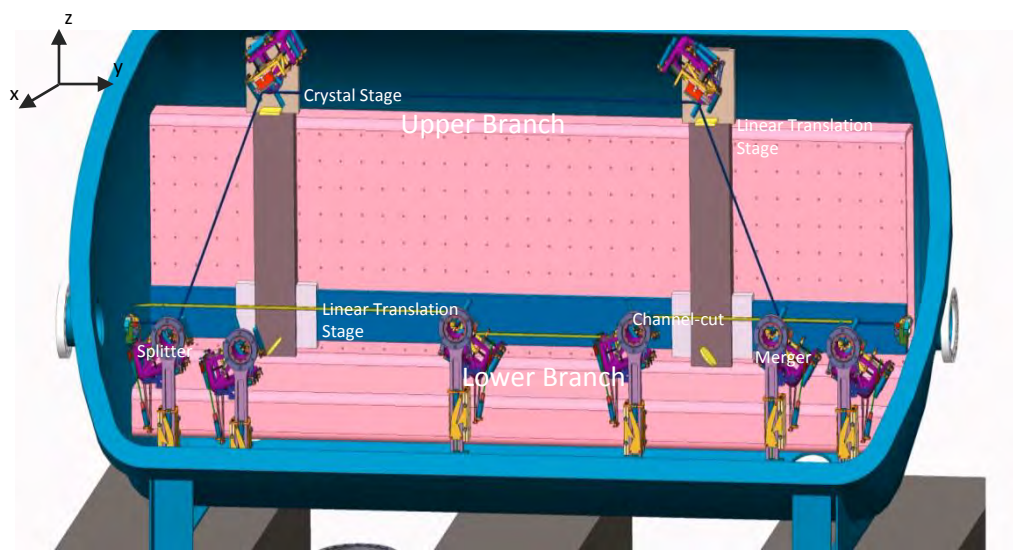

Figure 3-1: Mechanical design of upper and lower branch in split and delay line.
Provided by Dr. Tino Noll from TU Berlin.

As one can see from figure 3-1, the rails for the upper branch translation stages are mounted along the vertical extension of L shape optical bench. By controlling both vertical and horizontal linear translation stages, the upper branch crystals can move along z and y direction. This brings the requirements of maximum straightness and stiffness for the rails[4].

Besides, the splitters, channel cuts and mergers are mounted on the lower branch. There are two versions of splitter and merger for both thin and thick crystal. These should be installed along the opposite z direction as close as possible, and the thin crystal splitter or merger should precede the geometric splitter or merger. To avoid compromising on the maximum delay time, one can split the beam by using the thin crystal and continue in the inclined mode by using the thick crystal of the geometric merger. The empty spaces between splitter and first channel cut, and between second channel cut and merger, might host additional equipment, for example a CRL to enlarge the beam size of the pump beam or an attenuator[4].

By translations of the upper branch crystals, the total achievable time delay should be -10 to 800 ps, also in the inclined mode. Linear encoders on all the translations of the upper branch crystals are required[4].

## 3.2 Bragg Cage and Fine Alignment Stage

The Bragg rotation cage with rotary encoders provides the coarse adjustment for the Bragg angle of crystal, which was designed by Dr. Tino Noll and kept updating. A prototype design is presented in figure 3-2. There are two alignment stages, one is the coarse alignment stage with adjustable range of 18.8° ~ 40.2°, the other is the fine alignment stage with required accuracy of 0.1 µrad in pitch and 0.2 µrad in roll[4]. The Bragg cage will be installed on the vertical translation stage with 10 motors inside. Now the prototype is already finished in TU Berlin. Later it will be added to our test-stand for testing.

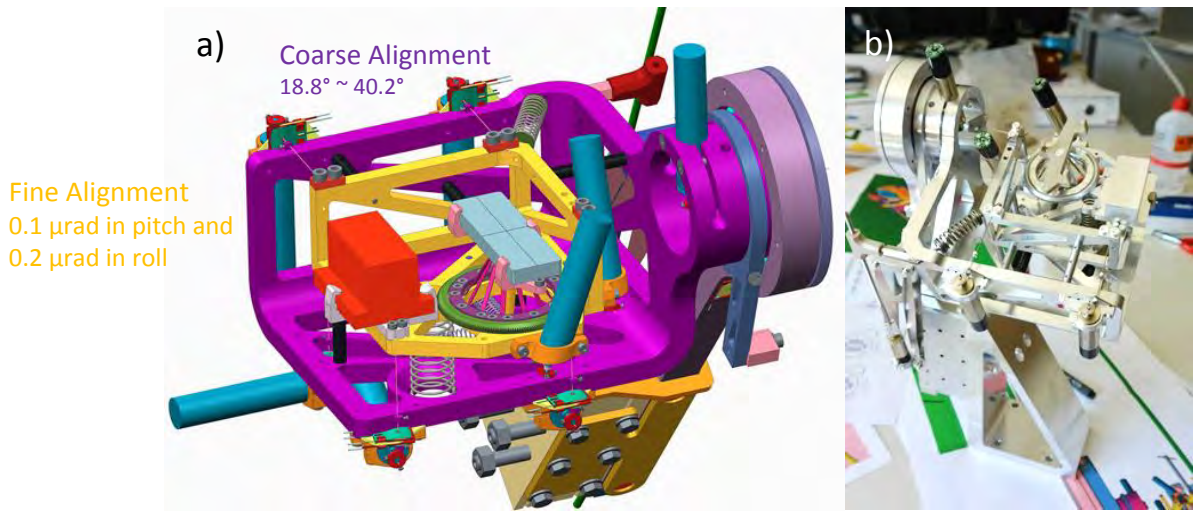

Figure 3-2: Mechanical design[a] and prototype[b] of Bragg cage and fine alignment stage.
Provided by Dr. Tino Noll from TU Berlin.

## 3.3  Laser Interferometer

In order to measure the angular variation of the fine alignment stage mentioned in Ch. 3.2, a laser interferometer will be employed here in SDL. The principle of laser interferometer is based on the light interference. The laser is separated into two beams and travel along different optical path as shown in figure 3-3, which leads to laser interference when the beams meet again. Also there is another reference signal from the original laser as comparison[4]. The displacement of the target can be obtained from the interference signal and the angular information can be deduced from two axes. For measuring the pitch and roll angles of the fine alignment stage, a 3-axis laser interferometer has been designed and the prototype are presented in figure 3-4.

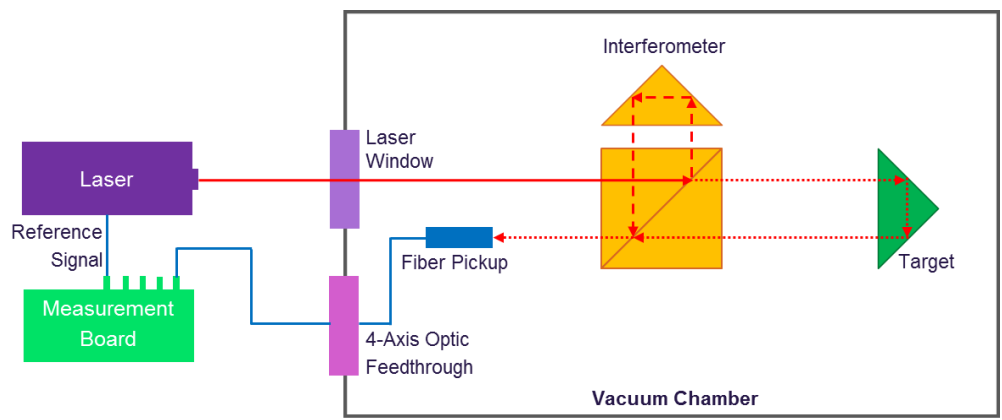

Figure 3-3: The schematic of laser interferometer in single axis.

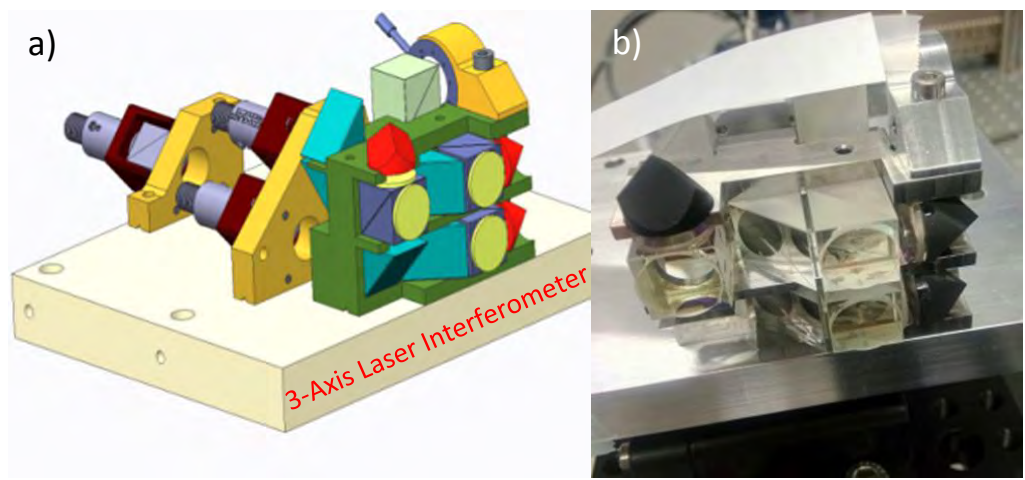

Figure 3-4: Mechanical design[a] and prototype[b] of the 3-axis laser interferometer.
Provided by Dr. Wei Lu.

## 3.4 Vacuum Vessel and Support Structure

The current designs of vacuum vessel and granite support structure are presented in the following figure 3-5.

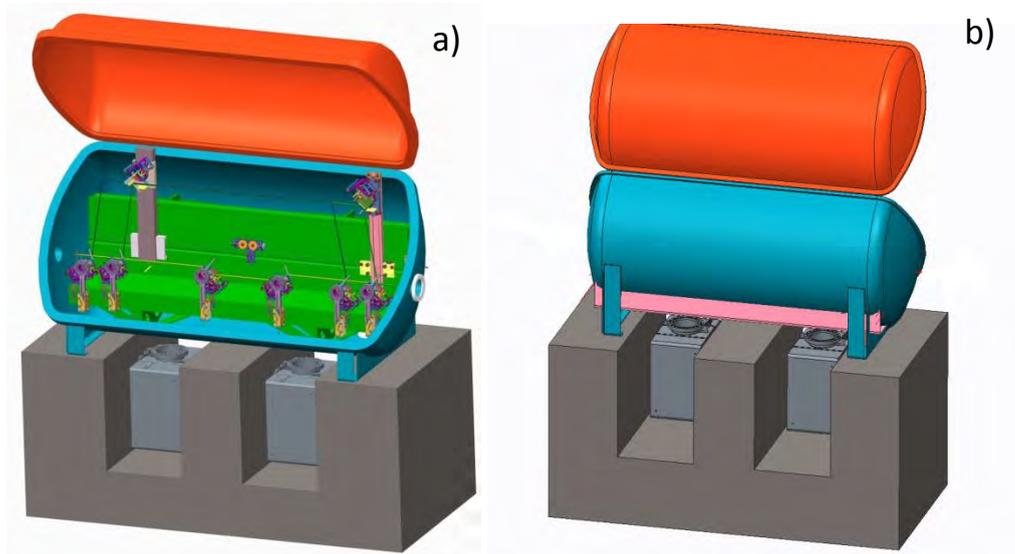

Figure 3-5: The front[a] and back[b] view of vacuum vessel and support structure of SDL.
Provided by Dr. Tino Noll from TU Berlin.

To avoid vibration during operation, the vacuum will be maintained only by two Agilent Star Cell 500 ion pumps, which are suspended below the vessel, filling empty spaces in the supporting granite. High vacuum ($10^{-6}$ mbar) can be achieved by a Pfeiffer HiPace 800 turbo pump, installing on a DN 200 port and a DN200 gate valve. An Edwards nXDS 15 scroll pump provides roughing vacuum ($10^{-2}$ mbar) for the turbo pump[4].

# 4 Mechanical Design of the Test-Stand for SDL

The aim of the test-stand for SDL is to partially simulate and implement the functions of SDL. By using this test-stand, the upper branch stage motion, electronic and optic configuration, vacuum degree and etc. can be tested. In this chapter, the designs I've done for the test-stand are presented.

## 4.1 Motion Design

In the test-stand, the upper branch motion of SDL is simulated. Since the motion of upper branch's left and right part are symmetric, to simplify this case, only half of the upper branch is taken into the consideration of test-stand design as shown in figure 4-1.

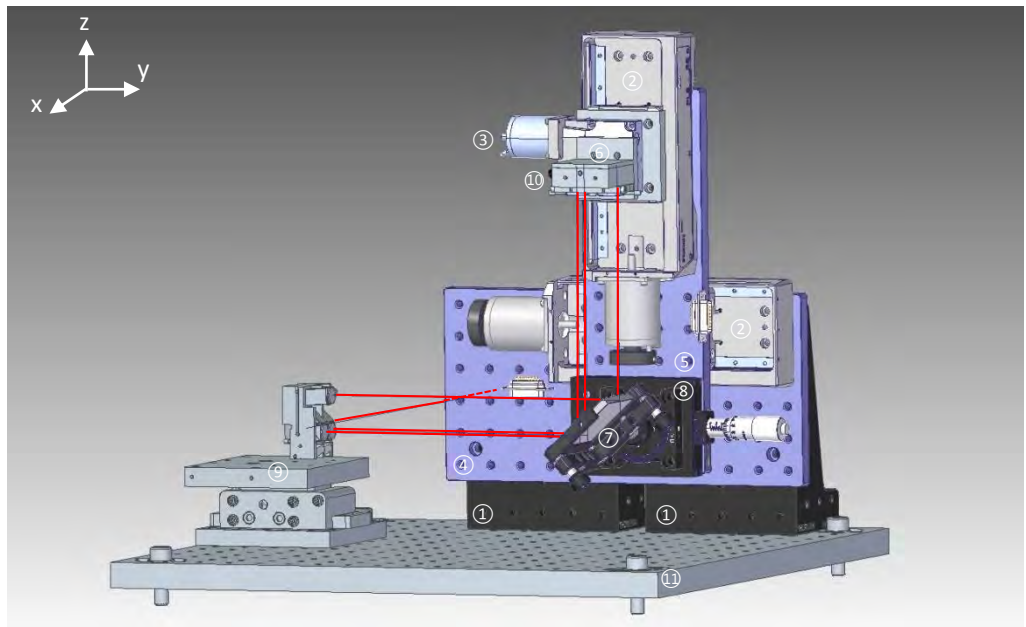

Figure 4-1: Mechanical design of upper branch for the test-stand.
Components List:
1. Large Angle Bracket, 1/4"-20 Holes, AP90RL, ThorLabs (×2)
2. Precision Linear Stage, PLS-85, PiMicos (×2)
3. Rotation Stage, RS-40, PiMicos (×1)
4. 150 mm x 300 mm - Aluminum Breadboard, MB1530F/M, M6 Taps, ThorLabs (×1)
5. 100 mm x 300 mm - Aluminum Breadboard, MB1030/M, , M6 Taps, ThorLabs (×1)
6. Joint Aluminum Breadboard, Manufactured in DESY HASYLAB (×1)
7. Reflection Mirror (×1)
8. Manual Rotational Stage (×1)
9. Interferometer (×1)
10. Retro-reflector (×1)
11. Bottom of Vacuum Chamber (×1)

Two large angle brackets (1) fixed on the bottom of the vacuum chamber (11) are the fundamental support for the translation stages (2). And a modified 150mm×300mm breadboard (4) with 12 M6 thread holes (C.3) are installed on that. Then a translation stage (2) is horizontally mounted on that, which provides the linear motion along y direction. The same way for the installation of vertical translation stage (2) along z direction. Another modified 100mm×300mm breadboard (5) with 8 M4 thread holes (C.3) is used.

A rotation motor (3) is installed on the vertical translation stage (2), which is necessary to operate the laser interferometer (9). Because of non-compatible thread holes and drill holes on rotation motor and retro-reflector (10), a T-shape aluminum breadboard (6) (C.3) was designed to make sure that the rotation axis of the retro-reflector is in the rotation center of the motor.

The incoming laser is along x direction and passes through the interferometer (9). Then it is separated into three beams and all of them propagate along y direction. After being reflected by the 45° angle mirror (7), the beams will propagate parallelly towards the z direction and reach the retro-reflector.

Figure 4-1 was drawn by Solid Edge software[8] using imported stepfiles and the assembly function under ordered mode.

## 4.2  Vacuum Chamber and Support Frame

The designs of vacuum chamber and support frame are shown in figure 4-2. The size of the vacuum chamber is 526mm×513mm×520mm and the support frame is 724mm×650mm×740mm.

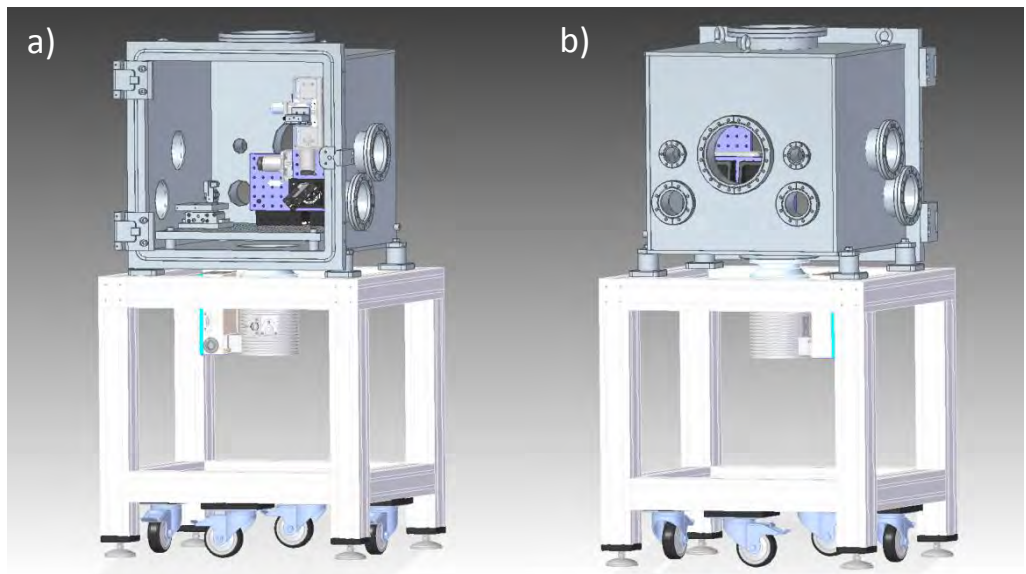

Figure 4-2: Front[a] and back[b] view of vacuum chamber and support frame for the test-stand.

The vacuum chamber is a steel cube with the front gate made by Acryglass. It is manufactured by Pfeiffer Vacuum GmbH for high vacuum usage. Thus it has well air tightness to achieve the degree of vacuum less than $10^{-6}$ mbar. There are two DN100 flange holes on both left and right side for view ports, electronic and optic feedthroughs. Other 6 flange holes on the back and top side are reserved for incoming laser, vacuum gauge and etc. Besides, there is a DN 160 flange hole on the bottom for the installation of turbo pump.

Regarding the support frame, it is an aluminum frame with four wheels for easy moving. The vacuum chamber is firmly fixed on this frame. Thus it should be strong and steady enough to support the heavy vacuum chamber and its inside components.

## 4.3 Electronic Configuration

In total there are 13 motors in the test-stand. Motor No. 1 to 10 are in the Bragg cage and alignment stage, and Motor A, B and C are the three translation and rotation stages in upper branch. That means it should have 13 D-Sub connectors in the electronic configuration. The important consideration for electronic configuration are the number and scale of flanges that will be used. The designs of both Plan A and B are aimed to connect the D-Sub feedthroughs to the 9 pins D-Sub connectors on motor side. Here is the detail configuration of each plan.

### 4.3.1 Electronic Configuration - Plan A

Provider: Allectra GmbH

Feedthroughs: 4×25 Pins, D-Sub, Male (×1);  4×9 Pins, D-Sub, Male (×1);

Connectors: 9 Pins, D-Sub, M (×10); 25 Pins, D-Sub, F (×4); 9 Pins, D-Sub, F (×15).

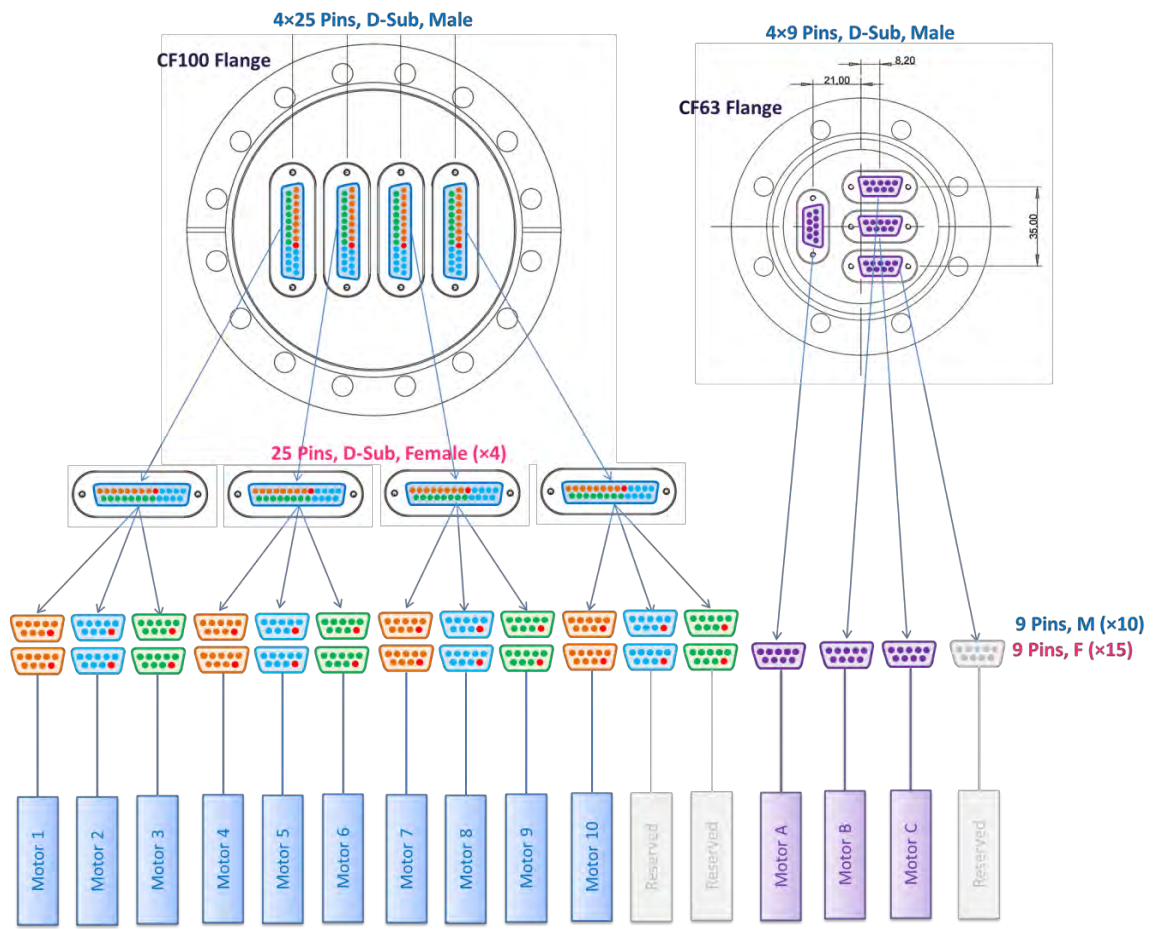

Figure 4-3: Schematic of electronic configuration for plan A.

For plan A, it is also need to specially indicate the pin connection between 25 pins and 9 pins D-Sub connector as shown in figure 4-4.

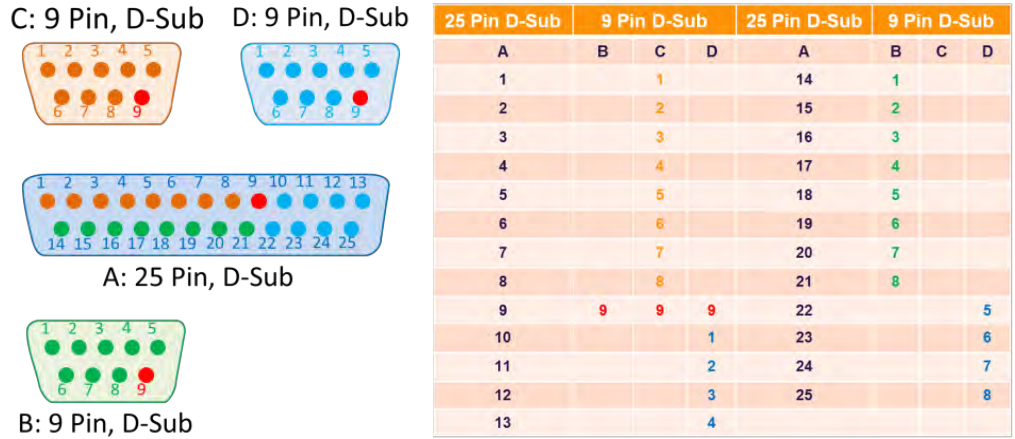

Figure 4-4: Schematic of pin connection between 25 pins and 9 pins D-Sub connector.

### 4.3.2 Electronic Configuration - Plan B

Provider: Allectra GmbH
Feedthroughs: 8×9 Pins, D-Sub, Male (×2);
Connectors: 9 Pins, D-Sub, F (×13).

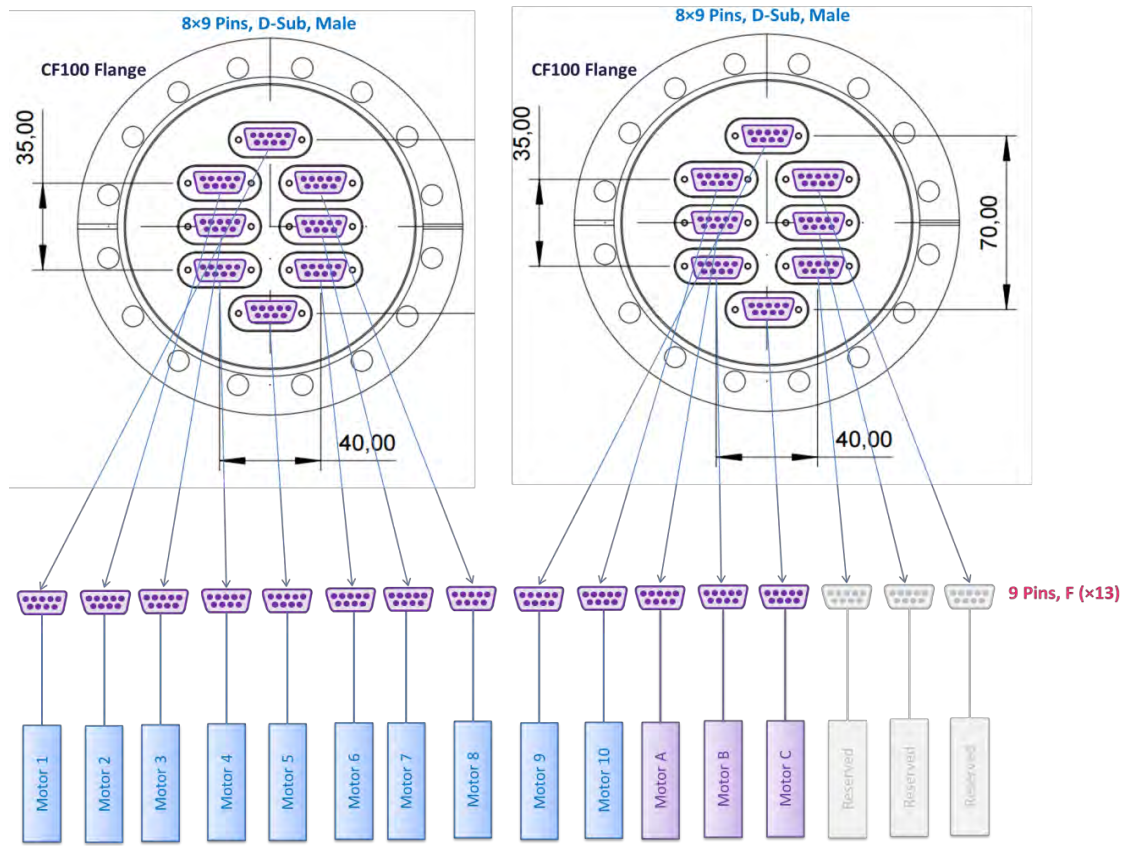

Figure 4-5: Schematic of electronic configuration for plan B.

The advantage of plan A (with one DN100 CF and one DN63 CF flange) is saving place by using smaller flange. While for plan B (with two DN100 CF flanges) the advantage is the simple configuration which is easy to handle.

# 5 Preparation and Test of the Test-Stand for SDL

This chapter describes the detail assembly procedure of the test-stand for SDL, the chamber vacuum test and the parasitic motion test.

## 5.1 Assembly of the Test-Stand for SDL

As presented design of the test-stand in Ch.4, the assembly of the test-stand is followed by below procedures with detail components listed.

| **List of Assembly Procedures and Detail Components** | | | | |
|---|---|---|---|---|
| **1.** | **Assembly of Chamber:** <br> The vacuum chamber[1.2] is fixed on the support frame[1.1]. Then nine blind flanges[1.3] and three view ports[1.4] are installed on the flange holes. | | | |
| 1.1 | **Support Frame** | 724mm×650mm×740mm, Al | ×1 | |
| 1.2 | **Vacuum Chamber** | 526mm×513mm×520mm, Steel | ×1 | Pfeiffer |
| 1.3 | **Blind Flange** | UHV DN 200 CF | ×1 | M6 Screw |
| | | UHV DN 100 CF | ×4 | M4 Screw |
| | | UHV DN 63 CF | ×2 | M3 Screw |
| | | UHV DN 40 CF | ×2 | M3 Screw |
| 1.4 | **View Port** | UHV DN 100 CF | ×2 | M4 Screw |
| | | UHV DN 160 CF | ×1 | M4 Screw |
| **2.** | **Assembly of Vacuum Test Components** <br> To test the vacuum, a turbo pump[2.5] is installed on the bottom of the chamber with pre-pump[2.6] connected to turbo pump by KF tubes[2.3]. A gate valve[2.4] is also installed on the turbo pump. On the back side of the chamber, a vacuum gauge[2.1] connecting with a pressure indicator[2.2] is installed. | | | |
| 2.1 | Vacuum Gauge | UHV DN 40 CF | ×1 | Pfeiffer |
| 2.2 | Pressure Indicator | Connected with Gauge | ×1 | Pfeiffer |
| 2.3 | Connection Tube | Connected with Turbo Pump, KF | ×2 | |
| 2.4 | Gate Valve | Connect with Tube | ×1 | |
| 2.5 | Turbo Pump | Use DN100 – 160 CF Reducer | ×1 | œrlikon |
| 2.6 | Pre-Pump | Connect to Turbo Pump | ×1 | |

| 3. | **Assembly of Mechanical Parts** <br> These procedures are followed as the descriptions in chapter 4.1 | | | |
|---|---|---|---|---|
| 3.1 | Large Angle Bracket | AP90RL | ×2 | Thorlabs |
| 3.2 | Aluminum Breadboard | 150 mm x 300 mm | ×1 | Thorlabs |
| 3.3 | Precision Linear Stage | PLS-85 | ×1 | PiMicos |
| 3.4 | Aluminum Breadboard | 100 mm x 300 mm | ×1 | Thorlabs |
| 3.5 | Precision Linear Stage | PLS-85 | ×1 | PiMicos |
| 3.6 | Rotation Stage | RS-40 | ×1 | PiMicos |
| 3.7 | Joint Al Breadboard | T-shape | ×1 | |
| 3.8 | Retro-reflector | 3 axes | ×1 | |
| 3.9 | Manual Rotation Stage | For Mirror | ×1 | |
| 3.10 | Reflection Mirror | | ×1 | |
| 3.11 | Interferometer | | ×1 | |
| 4. | **Assembly of Electronic Parts** <br> The electronic feedthroughs[4.1] are installed by replacing the blind flanges on the chamber. Then the motor are connected with it by using D-Sub connectors[4.2]. | | | |
| 4.1 | Electronic Feedthrough | D-Sub Male, DN 100, 8 × 9 pin | ×2 | Allectra |
| 4.2 | D-Sub F. Connector | D-Sub Female, 9 pin | ×13 | Allectra |
| 5. | **Optical Parts** <br> The optic feedthrough[5.1] is also installed by the same way as the electronic feedthrough. And the optic fibers[5.2] are connected with the optic feedthroughs. | | | |
| 5.1 | Optic Feedthrough | CF 100, 4 axes | ×1 | |
| 5.2 | Optic Fiber | | | |

Table 5-1: List of assembly procedures and detail components of the test-stand for SDL.

Following the above procedures, finally I completed the assembly of the test-stand as shown in figure 5-1. The left side is the overview of support frame and chamber with mechanical components inside. And the detail constructions of mechanical components are shown in the right figure.

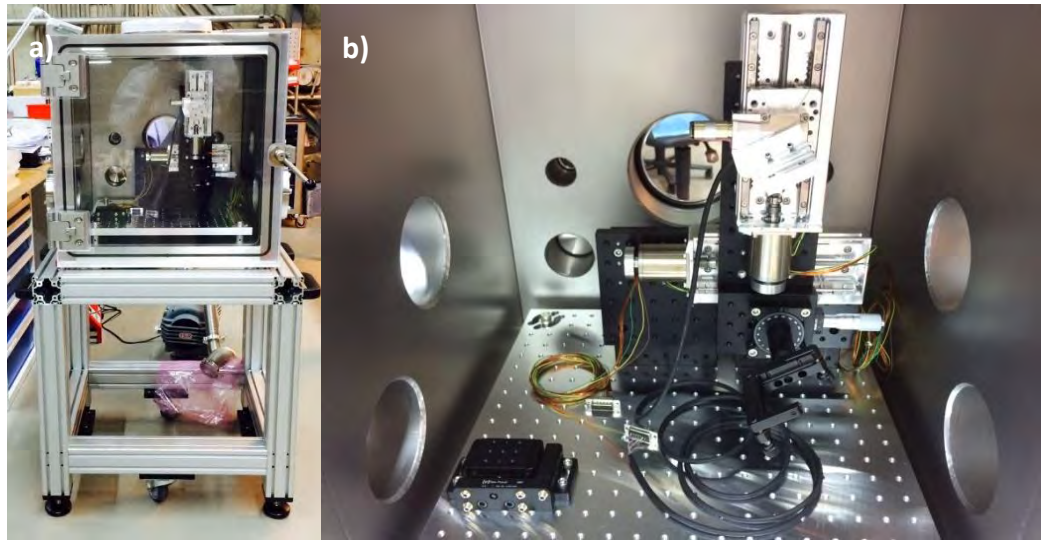

Figure 5-1: Construction of the test-stand for SDL in Hera Süd.

## 5.2 Chamber Vacuum Test

The vacuum test of the chamber was done in Hera Süd at MID station. The turbo molecular pump used for test can achieve high vacuum below $10^{-6}$ mbar. To test the vacuum, it is needed to assemble the gauge and turbo pump on the chamber as described in table 5-1. And the pre-pump is connected with turbo molecular pump as shown in figure 5-2.

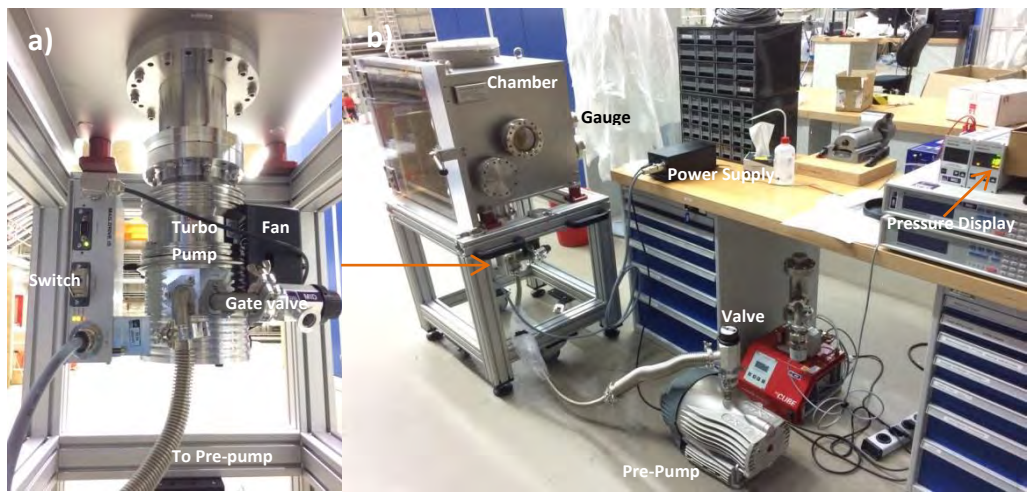

Figure 5-2: Configuration of chamber vacuum test in Hera Süd.

In the test experiment, the pre-pump is started with ambient pressure. To avoid the damage of turbo caused by big difference of pressure on both sides in the beginning, the turbo pump is only switched on after the pressure reaches $10^{-1}$ mbar, and runs for enough time to achieve $10^{-6}$ mbar. The recorded pressure data versus time are drawn in figure 5-3 with linear scale on the left and logarithmic scale on the right. The operation procedures are also indicated in figure 5-3 for clearly illustration.

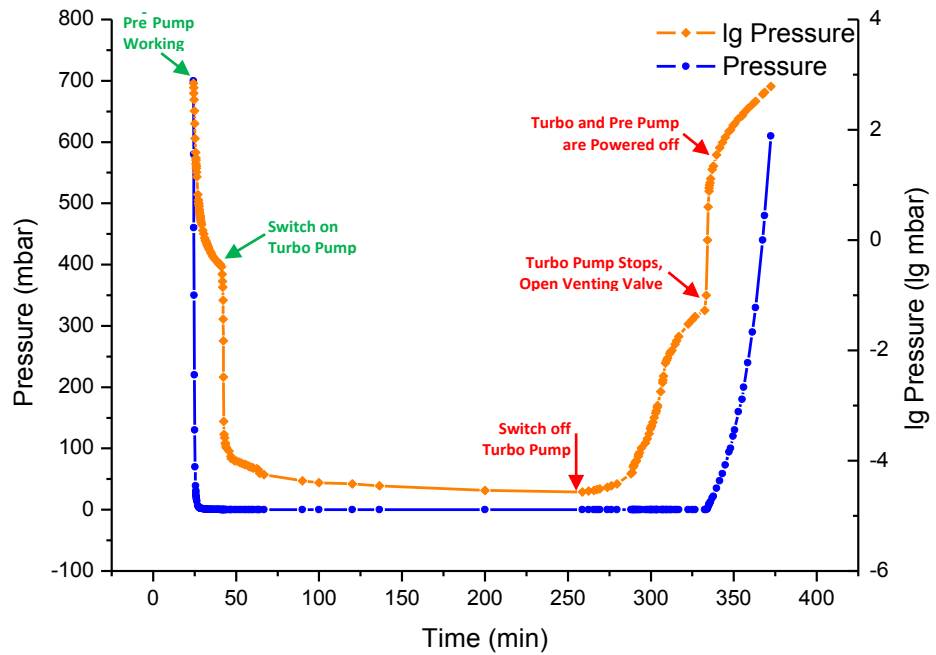

Figure 5-3: Vacuum test curves of original vacuum pressure and log 10 pressure versus time.

| Time | Pressure | Pump Operation | Light Indicator |
|---|---|---|---|
| (min) | mbar | On/Off and State | On Turbo Pump |
| 0 | 1000 | Switch on Pre-pump | N/A |
| 40 | $3.5 \times 10^{-1}$ | Switch on Turbo Pump | Green and Yellow on |
| 258 | $2.7 \times 10^{-5}$ | Switch off Turbo Pump | Green Flash / Yellow on |
| 312 | $1.3 \times 10^{-2}$ | Turbo Pump <200 Hz | Green and Yellow Flash |
| 332 | $5.3 \times 10^{-2}$ | Turbo Pump stops | Green off / Yellow on |
| 335 | 8.8 | Turbo Pump power off | Green off / Yellow off |
| 341 | 47 | Pre-pump power off | N/A |

Table 5-2: Key operations of pumps and their corresponding pressure, time and light indicator

With the logarithmic scale data, the pressure changes during the pumping process can be illustrated. The achieved pressure by pre-pump is around $10^{-1}$ mbar while with switching on turbo pump, the pressure can go down to $2.7 \times 10^{-5}$ mbar as listed in table 5-2. It is expected to reach $10^{-6}$ mbar by longer time pumping.

## 5.3 Cable Fabrication and Parasitic Motion Test

The tests of two translation motors and one rotation motor are operated through the terminal in Hera Süd, which requires the standard Lemo connector. Thus I fabricated three cables in XFEL's work shop for the testing.

### 5.3.1 D-Sub - Lemo Cable Fabrication

The XFEL standard Lemo connector includes 8 crimp contacts, plastic housing, and metal shell. To fabricate this kind of Lemo connector, firstly use a 6 meters long cable and remove 25mm of the rubber cover according to the cable stripping lengths (C.4). Inside the cable there are 8 copper wires with different color marks. Remove 4mm of 8 copper wires' cover and install the metal crimp contact for each of them. Following by that, insert 8 contacts to a plastic housing in color order as shown in figure 5-5 and assemble the metal shell.

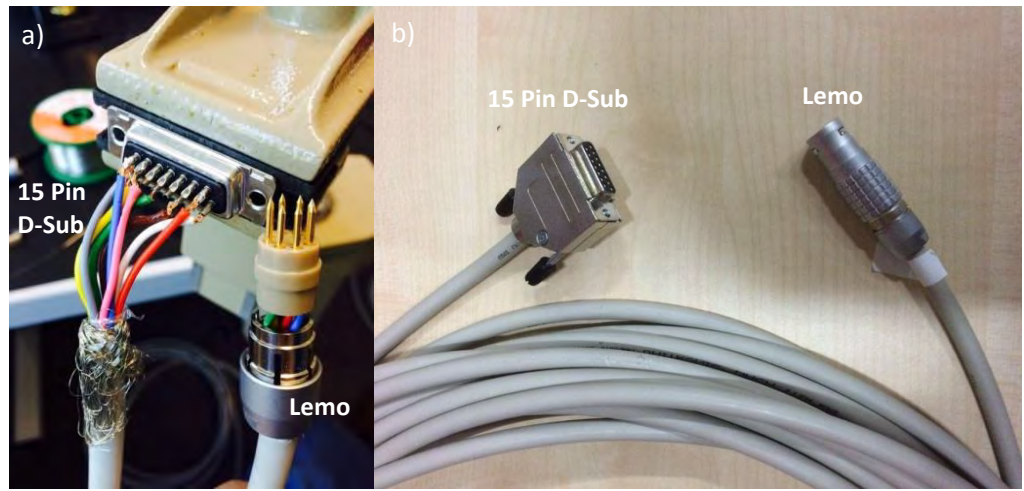

Figure 5-4: Inside[a] and outside[b] view of Lemo and D-Sub Connector for translation stage.

Regarding the D-Sub connector side, the electronic configuration is different for translation and rotation stage motors. The 8 copper wires are connected with 15 pins female D-Sub using a welder and fusible alloy solder. The detail electronic schematics for both of them are shown in figure 5-5 and figure 5-6 respectively.

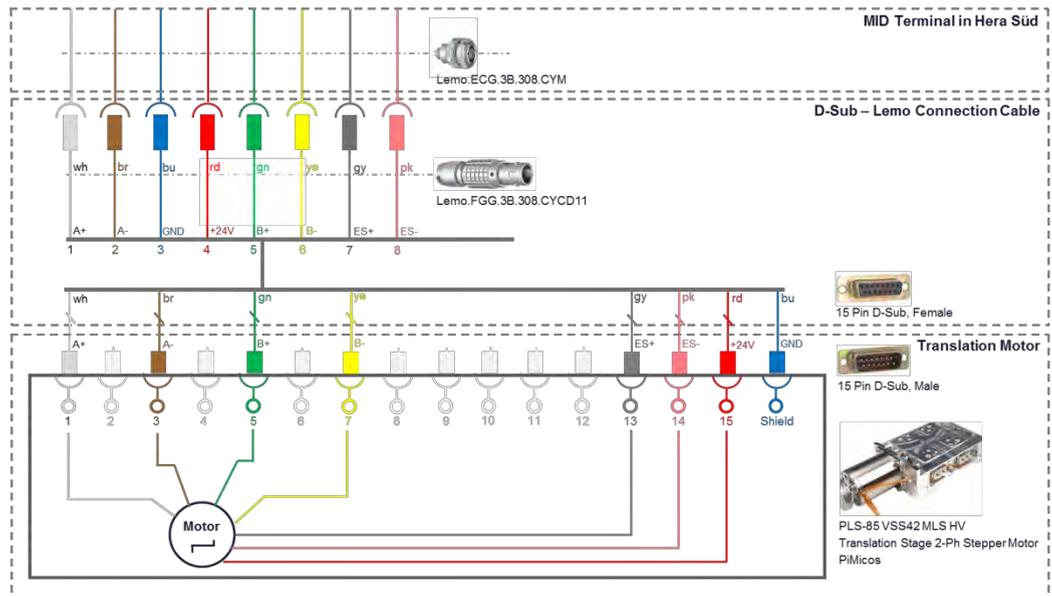

Figure 5-5: The electronic connection schematic for translation stepper motor.

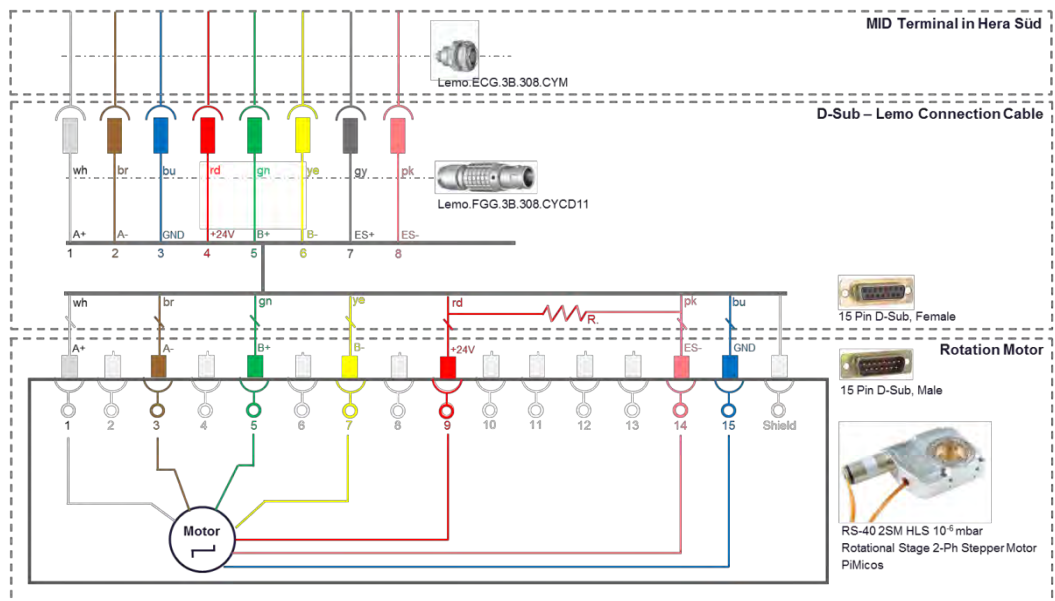

Figure 5-6: The electronic connection schematic for rotational stepper motor.

In the electronic connection schematic of rotational stepper motor, the end switch plus is not connected because of continuous rotation. According to XFEL stepper motor connection in chapter C.4, for rotational stepper motor, there should be a resistance between Pin No. 9 (applied voltage) and Pin No.14 (end switch minus).

### 5.3.2 Karabo[13] Software Controlling

The European XFEL's controlling system is Karabo software framework, which combines control, data management and scientific computing tasks. It is currently under development and used in the controlling of parasitic motion test. Here I present the Karabo Graphical User Interface (Karabo GUI) for motor testing.

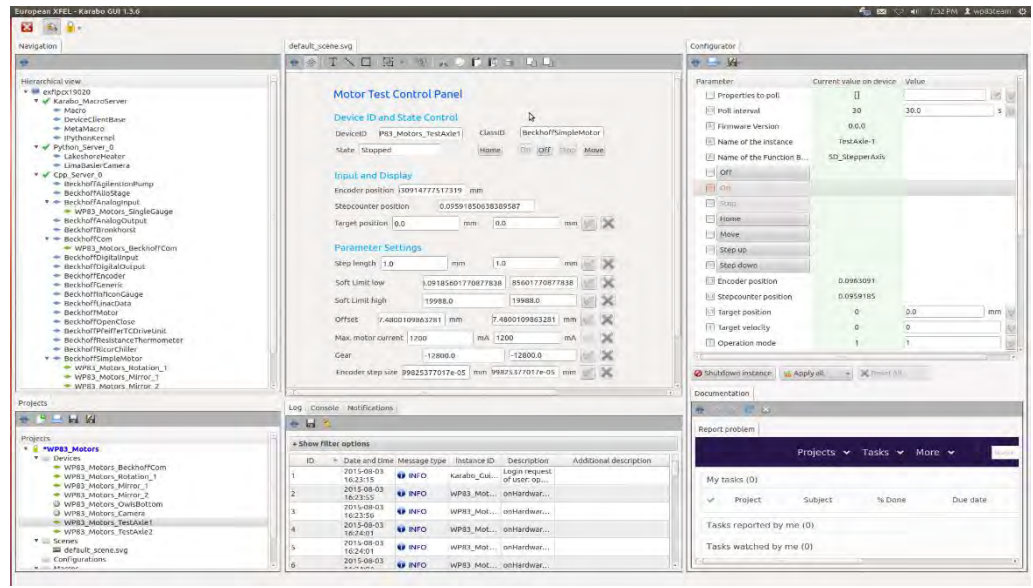

Figure 5-7: Karabo GUI 1.3.6-based interface with developed control panel for translation motor test[13].

After connected the motor and controller with D-Sub – Lemo cable, run the Karabo GUI 1.3.6 in MID terminal, select the corresponding motor test axle and set the parameters in the configuration section on the right side. For translation and rotation motors, the technical parameters are listed in the following table 5-3.

| Motor Type | Translation | Rotation |
|---|---|---|
| Phase Current | 1.2A | 0.25A |
| Step Angle | 1.8° | 15° |
| Steps | 200 | 24 |
| Coil-Resistance | 1.6Ω | 12.5 Ω |
| Coil-Inductance | 3 mH | 5.5 mH |
| Holing-Torque | 130 mNm | 6 mNm |
| Pitch | 1mm/rev | 4°/ turn (gearbox) |
| Gear Ratio | — | 387283/5103=75.892187 |
| Resolution/Fullstep | 5µm | 0.0021961° |

Table 5-3: Technical Parameters of translation and rotational motors.

To set the configuration in Karabo GUI, the settings according to the calculation based on the parameters of motors are needed, as shown in table 5-4:

| Parameter | Current Value on Device | Need to be set |
|---|---|---|
| Compatibility | 1.0 | |
| ClassID | BeckhoffSimpleMotor | |
| ServerID | Cpp_Server_0 | |
| DeviceID | WP83_Motors_TestAxle1 | |
| Archive | True | ☑ |
| Use Timeserver | False | |
| Progress | 0 | |
| State | On/Off/Stopped/Running/Idle | |
| Performance Statistics | — | |
| Beckhoff Communication | WP83_Motors_BeckhoffCom | |
| PLC ID | 0x70201010 | |
| On Autonomous Hardware | followSilently | ☑ |
| Hardware state (hex) | 0x1AB220 | |
| Hardware state (bits) | 0000 0000 0001 1010 1011 0010 | |
| Hardware state (string) | Normal_Stopped | |
| Hardware Err description | 0x0 | |
| **Request Hardware Values** | | Button |
| **Safe** | | Button |
| **Normal** | | Button |
| **Override** | | Button |
| Properties to poll | [] | ☑ |
| Poll interval | 30 | ☑ |
| Firmware Version | 0.0.0 | |
| Name of the instance | TestAxle-1 | |
| Name of the Function | SD_StepperAxis | |
| **Off** | | Button |
| **On** | | Button |
| **Stop** | | Button |
| **Home** | | Button |
| **Move** | | Button |
| **Step up** | | Button |
| **Step down** | | Button |
| Encoder Position | 0.0963091 | |
| Stepcounter Position | 0.0959185 | |
| Target Position | 0 | ☑ |
| Target Velocity | 0 | ☑ |
| Operation Mode | 1 | |
| Step Length | 1.0mm | ☑ |
| Stopped/Idle on target | False | ☑ |
| At h/w limit high | False | |
| At h/w limit low | True | |
| At s/w limit high | False | |
| At s/w limit low | True | |
| Master Slave Correlation | 1 | ☑ |
| Gear[a] | -12800 | ☑ |
| Encoder step size[b] | -7.8125E-05mm | ☑ |
| Controller accuracy | 0.00125mm | ☑ |
| Controller constant | 0.8 | ☑ |
| Max. acceleration | 500 | ☑ |
| vMax | 32767 | ☑ |
| Max. motor current[c] | 1200 | ☑ |
| Soft Limit low[d] | -20000 | ☑ |
| Soft Limit high[e] | 20000 | ☑ |
| Invert limits | — | ☑ |
| Offset | -357.48 | ☑ |
| Is internal Counter | True | |
| Reduced current | 4 | ☑ |
| Epsilon | 0.01 | ☑ |
| Backlash | 1 | ☑ |
| Number of PLC Cycles | 1 | ☑ |
| Is limitless | False | ☑ |
| Save h/w limit | True | ☑ |
| Check limit consistency | True | ☑ |

Table 5-4: Parameter settings of translation stage motor in Karabo GUI 1.3.6.
[a]Gear =64 microsteps per step × step number=64×200=12800;
[b]Encoder step size=1/Gear=1/12800=7.8125E-05;
[c]Max.motor current: according to table 5-3;
[d]Soft limit low: The low limitation of target position;
[e]Soft limit high: The high limitation of target position.

Now the preparation work of controlling system is done. Parameters can be set and programming code can be written for the desired procedure like motor scan. The controlling of motors can be easily integrated and operated in the control panel as shown in the centre of figure 5-7.

### 5.3.3  Parasitic Motion Test

**Parasitic Motion Test along X-direction**

In order to test the Parasitic Motion of the motor located along x-direction, I developed the mechanical configuration as shown in figure 5-8. Motor 1 is used for the motion along x, and two digital indicators purchased from Studenroth Präzisionstechnik GmbH are used to measure the parasitic motion on both sides. The distance between the left and right indicator is 210 mm. Other breadboards are employed for conjunction and construction.

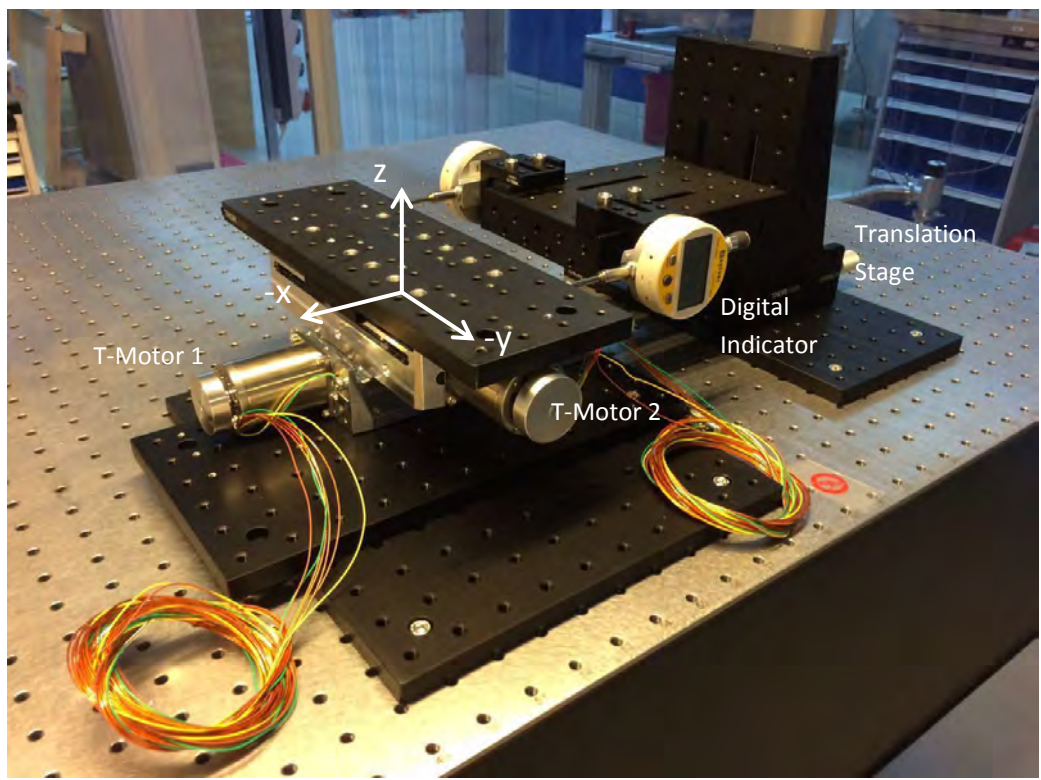

Figure 5-8: Configuration of parasitic motion test for two translation motors.

After the setting up of parasitic motion test, the motors are connected to MID terminal using fabricated D-Sub - Lemo cables. Then start the Karabo GUI 1.3.6, select the Test-Axle1 (connected with T-Motor 1) and click "on" and "home" button to reset the encoder position. Set the position of T-Motor2 to 50 mm and move the indicator to touch the edge of the breadboard fixed on the T-Motor2, and reset both indicators to zero. Since the digital indicators' measurement range (26 mm) is not enough for the whole stage travel length (100 mm), I have to divide this measurement into 5 parts with each of them overlapping for 5 points. The values of both left and right indicator are noted down as a function of motor position.

The angular change is calculated by the following formula.

$$Angular\ Change\ (\mu rad) = arctan\left[\frac{indicator\ right\ (mm) - indicator\ left\ (mm)}{210 mm}\right] \times 10^6$$

The angular change versus motor position is drawn in figure 5-9.

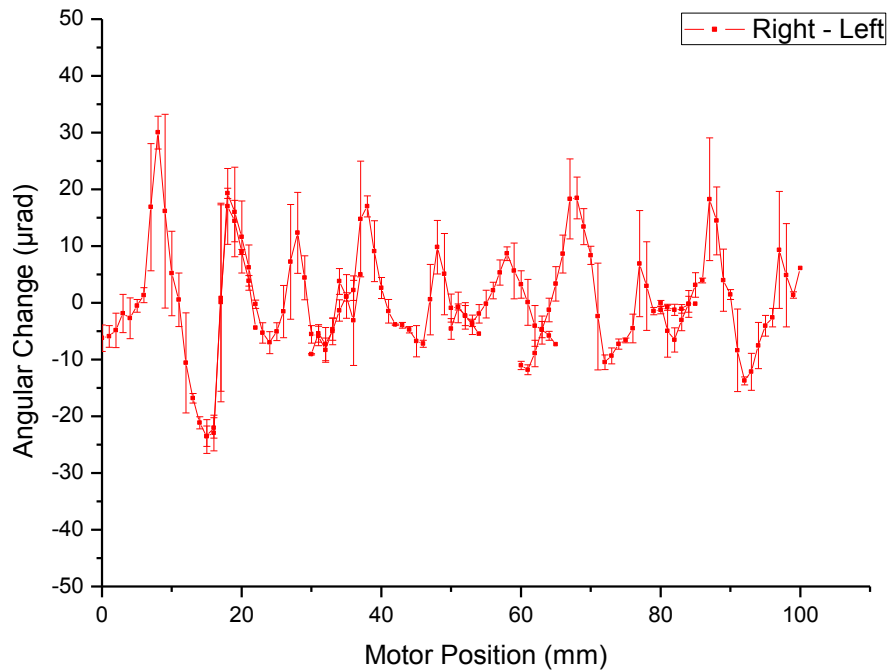

Figure 5-9: Parasitic motion of T-Motor 1 along x direction.

As one can see in figure 5-9, the angular change is within the range of ±35 μrad, and the calculated standard deviation of σ=9.11462 μrad is obtained, which is much better than the given value of ±100 μrad in product technical sheet (C.5).

The possible interpretations for the periodic behaviour shown in figure 5-9 are the mechanical errors of the stage induced by manufacturing process and mechanical design. Moreover, the measurement with mechanical contacts could also bring uncertainty and errors to the results. Thus the non-contact laser interferometry method should be a better option for this type of test for its higher accuracy and smaller perturbation. While the result here could be served as a reference for comparison.

**Parasitic Motion Test along X and Y Direction**

This is aimed to test the parasitic motion along Y direction with the motor along x direction also moving step by step. Since the measure instrument is mechanical indicator, it could bring uncertainty during the measurement. Thus the test cannot be performed with current instrument while the efficient and precise way to perform this test is laser interferometry. However the laser system is still not delivered. Nevertheless, the programming code based on python was developed for this parasitic motion test in Karabo GUI[13] 1.3.6. The motor scan and meander motion procedure were defined in the following code.

```python
>>> from Karabo.api import *  # import everything important for macros

class motorScan(Macro):
    motor1Id = String(defaultValue='WP83_Motors_TestAxle1')
    motor2Id = String(defaultValue='WP83_Motors_TestAxle2')
    stepLength = Float(defaultValue=1)
    nStepsX = Int(defaultValue=3)
    nStepsY = Int(defaultValue=3)
    @Slot()
    def hello(self):
        print("Hello world!")

    @Slot()
    def scanMotor2(self):
        for i in range(self.nStepsX):
            print("step #{}".format(i))
            self.stepUp(self.motor2Id)
    @Slot()
    def scanMotor1(self):
        for i in range(self.nStepsX):
            print("step #{}".format(i))
            self.stepUp(self.motor1Id)

    @Slot()
    def meander(self):
        with getDevice(self.motor1Id) as motor1:
            with getDevice(self.motor2Id) as motor2:
                origin = [motor1.encoderPosition,motor2.encoderPosition]
        ySteps = 0
        while(ySteps < self.nStepsY):
            for i in range(self.nStepsX):
                print("step #{}".format(i))
                self.stepUp(self.motor2Id)
            if(ySteps >= self.nStepsY):
                break
            self.stepUp(self.motor1Id)
            ySteps = ySteps + 1
```

```python
        for i in range(self.nStepsX):
            print("step #{}".format(i))
                self.stepDown(self.motor2Id)
            if(ySteps >= self.nStepsY):
                break
            self.stepUp(self.motor1Id)
            ySteps = ySteps + 1

        with getDevice(self.motor1Id) as motor1:
            with getDevice(self.motor2Id) as motor2:
                motor1.targetPosition = origin[0]
                motor2.targetPosition = origin[1]
                waitUntil(lambda: motor1.state == "Idle")
                waitUntil(lambda: motor2.state == "Idle")

    @Slot()
    def moveToStepLength(self):
        with getDevice(self.motor1Id) as motor1:
            with getDevice(self.motor2Id) as motor2:
                motor.targetPosition = self.stepLength
                motor.move()
                waitUntil(lambda: motor.state == "Idle")
                print("done moving")

    def stepUp(self, motorId):
        with getDevice(motorId) as motor:
            motor.targetPosition = motor.targetPosition + self.stepLength
            motor.move()
            waitUntil(lambda: motor.state == "Idle")
            print("done moving")

    def stepDown(self, motorId):
        with getDevice(motorId) as motor:
            motor.targetPosition = motor.targetPosition - self.stepLength
            motor.move()
            waitUntil(lambda: motor.state == "Idle")
            print("done moving")

#for the command line, instead of

        with getDevice(self.motorId) as motor:
            motor.targetPosition = self.stepLength
            motor.move()
            waitUntil(lambda: motor.state == "Idle")
            print("done moving")

#do:
        motor = connectDevice(deviceId)
        motor.targetPosition = self.stepLength
        motor.move()
        waitUntil(lambda: motor.state == "Idle")
        print("done moving")
```

## 5.4 Outlook and Discussion

The work I have done in this internship report is just the beginning of the test-stand for SDL, which is a small but important part for the magnificent European XFEL project. It's no doubt that three months are not enough to finish that, still there are a lot of following tests and modifications need to be done. Mainly I would like to summary the work I have done and the following arrangements for this test-stand. Also the existed problems I faced and the foreseeable challenges will be discussed here.

The prime work I was engaged in is design and construction of the test-stand for SDL. This part was almost completed except for the missing of interferometer and Bragg cage, which are still in processing by TU Berlin and MID group. Besides, the purchase of chiller and optic feedthrough is also under consideration. Regarding the construction of the test-stand, I finished the assembly of mechanical components. The following work that need to be done is assembly of interferometer, assembly of Bragg cage, arrangement of electronic and optic components, parasitic motion test along x and y, rotation motor test and etc.

During these design and construction processes, the problems I would like to point out are the vacuum component consideration and the test of parasitic motion. For the vacuum components, some of breadboards I used are not vacuum compatible, which need to be replaced later. Regarding the parasitic motion test, later the interferometer will be used because of its higher accuracy and less perturbation. The results could be compared with present test results.

The probably challenges for this test-stand project would be the accuracy controlling and the software controlling. To approach the very strict accuracy requirements for extremely intense and ultrashort X-ray flashes, it requires precise integration of every components and good function of each parts. Problems can be caused by any wrong operation and non-compatible part. Regarding the software controlling, since the Karabo GUI 1.3.6 is under upgrade and maintenance, there are some bugs when it runs for a certain time. It needs to be fixed with the help from Control & Analysis Software group at European XFEL.

It would be great to integrate the measurement output and the controlling input with the Karabo GUI and modify the control panel to assist the user to treat the data much more convenient and efficient.

# A  Acknowledgments

This internship was supported by European XFEL and DESY in Hamburg, Germany. I would like to thank my supervisor Dr. Wei Lu from MID group at European XFEL for his great support and guidance. I also would like to thank Dr. Anders Madsen as the group leader of MID at European XFEL who provided insight and expertise that greatly assisted my internship.

I express my warm thanks to Prof. Sergio Di Matteo, Prof. Philippe Rabiller and Mrs. Christiane Cloarec at Université de Rennes 1 for their continuous support, great help and suggestions during my internship.

In addition, I would like to thank Mr. Thomas Roth, Dr. Jörg Hallmann, Mr. Gabriele Ansaldi, Mr. Andreas Schmidt, Mrs. Birthe Kist, Dr. Alessandro Silenzi, Mr. Bernard Baranašić and Mr. Kai-Erik Ballak from European XFEL for their helps and comments that greatly improved the manuscript, also thank Ms. Mareike Fritz from HR department at European XFEL and Ms. Karin Kleinstück at TU München & LMU München for kindly helping me in application of German Visa.

Finally, an honorable mention goes to my family and relatives in Anqing China and my friends Zhang Lijuan in Xiamen; Peng Lan, Li Pujun and Liang Zhiqin in Beijing; Wu Fanglüe in Texas; Togzhan Nurmukanova in Rennes; Apoorva Ambarkar, Marta Mirolo and Asmaa Shaaban in Grenoble; Xiang Xie and Anjani K Mourya in Paris; Shambhavi Pratap in Lausanne; Sujeet Dutta in Hamburg, for their understandings and supports on me during this internship.

Thank you all,

August 1st 2015
Hamburg, Germany

Hong XU

Signature:

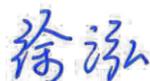

# B     Bibliography

# C     Appendix

Here I attached the indexes, drawings and data I used in this report for reference.

## C.1     Index of Figures

## C.2     Index of Tables

## C.3 Drawings of Modified Breadboards

**Modified Breadboard by adding 8×M4 Threads (in mm)**

**Modified Breadboard by adding 12×M4 Threads (in mm)**

Contact: Hong XU
TÉL: 6829

## C.3    Drawing of Joint Breadband

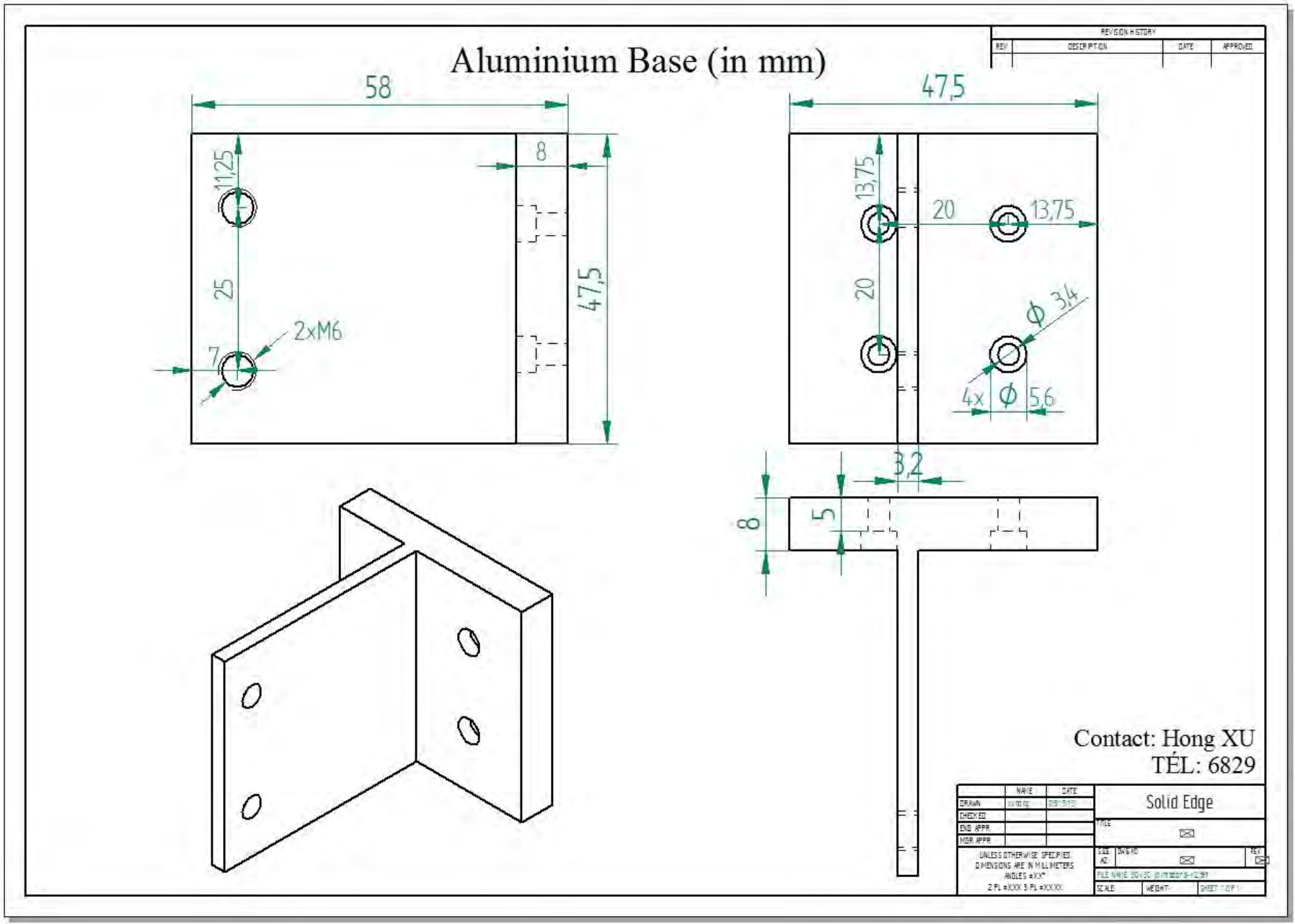

## C.4 Stepper Motor Connection

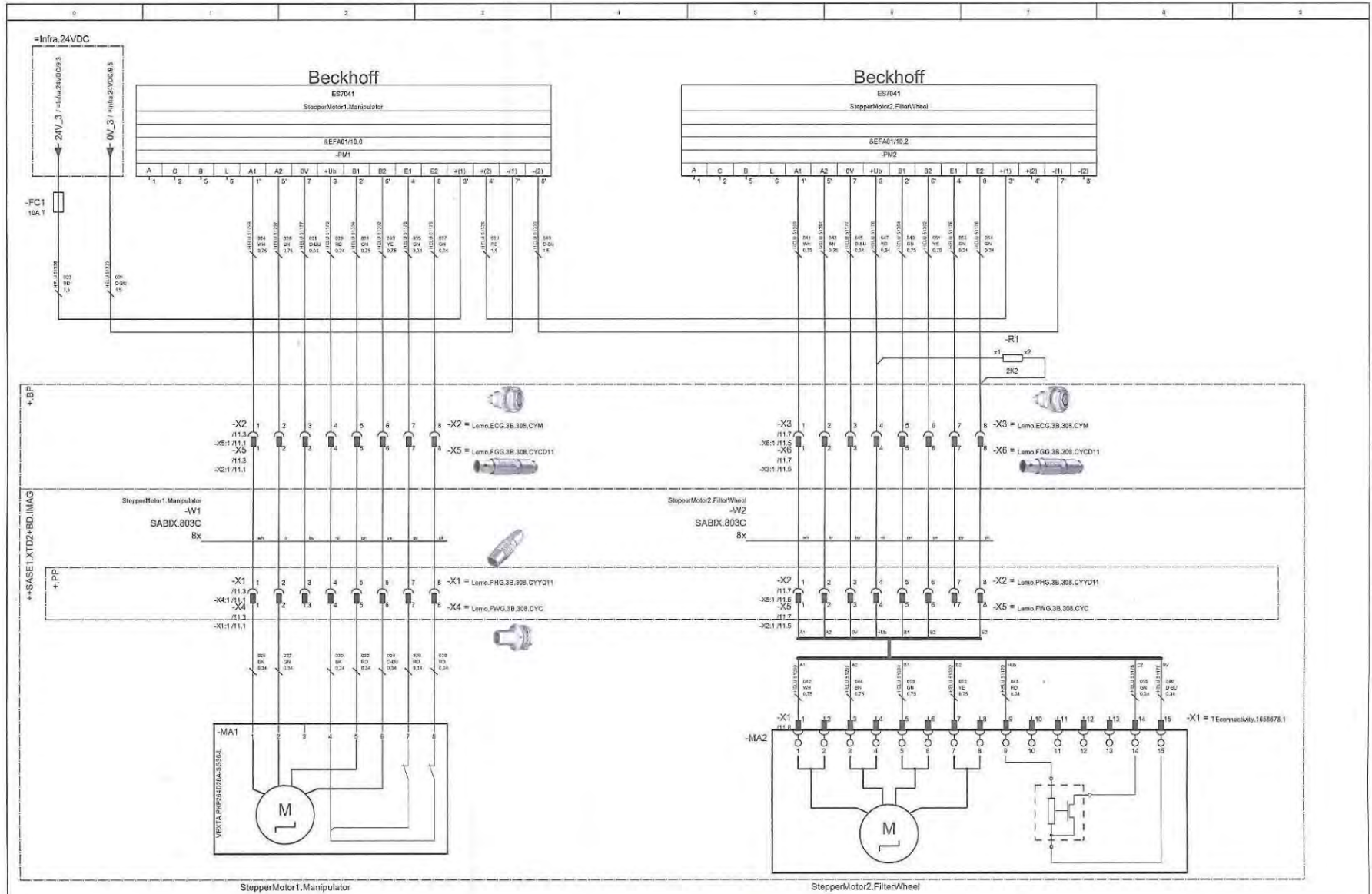

## C.4    Lemo Cable Stripping Lengths

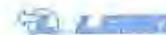

### Cable assembly (B, K, S and E series)

**Cable stripping lengths (B series)**

- M1  straight plugs and sockets with cable collet, clamping type D or M (solder or crimp contacts)
- M3  elbow plugs (90°) with cable collet, clamping type D or M (solder or crimp contacts)

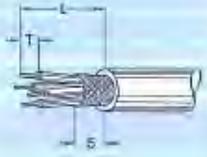

| Series | Type | ø contact A (mm) | M1 Solder L | M1 Solder S | M1 Solder T | M1 Crimp L | M1 Crimp S | M1 Crimp T | M3 Solder L | M3 Solder S | M3 Solder T | M3 Crimp L | M3 Crimp S | M3 Crimp T |
|---|---|---|---|---|---|---|---|---|---|---|---|---|---|---|
| 00 | 302/303/304 | 0.5 | 7.0 | 4 | 2.5 | 10.0 | 4 | 3.0 | 9.5 | 4 | 2.5 | 12.5 | 4 | 3.0 |
| 0B[1] | 302/303 | 0.9 | 14.5 | 7 | 3.5 | 17.0 | 7 | 4.0 | 19.5 | 7 | 3.5 | 22.0 | 7 | 4.0 |
| 0B[1] | 304/305 | 0.7 | 14.5 | 7 | 3.5 | 17.0 | 7 | 4.0 | 19.5 | 7 | 3.5 | 22.0 | 7 | 4.0 |
| 0B[1] | 306/307/309[2] | 0.5 | 14.0 | 7 | 2.5 | 18.0 | 7 | 3.0 | 19.0 | 7 | 2.5 | 23.0 | 7 | 3.0 |
| 1B[1] | 302/303 | 1.3 | 14.5 | 8 | 3.5 | 18.0 | 8 | 4.0 | 25.5 | 8 | 3.5 | 28.0 | 8 | 4.0 |
| 1B[1] | 304/305 | 0.9 | 14.5 | 8 | 3.0 | 18.0 | 8 | 4.0 | 25.5 | 8 | 3.0 | 28.0 | 8 | 4.0 |
| 1B[1] | 306/307/308 | 0.7 | 14.5 | 8 | 3.0 | 18.0 | 8 | 4.0 | 25.5 | 8 | 3.0 | 28.0 | 8 | 4.0 |
| 1B[1] | 310/314/316 | 0.5 | 16.5 | 8 | 2.5 | – | – | – | 27.5 | 8 | 2.5 | – | – | – |
| 2B | 302 | 2.0 | 19.0 | 9 | 4.0 | 22.0 | 9 | 5.5 | 30.0 | 9 | 4.0 | 33.0 | 9 | 5.5 |
| 2B | 303 | 1.6 | 19.0 | 9 | 3.5 | 22.0 | 9 | 5.5 | 30.0 | 9 | 3.5 | 33.0 | 9 | 5.5 |
| 2B | 304/305/306/307 | 1.3 | 18.0 | 9 | 3.5 | 20.0 | 9 | 4.0 | 29.0 | 9 | 3.5 | 31.0 | 9 | 4.0 |
| 2B | 308/310 | 0.9 | 17.0 | 9 | 3.0 | 20.0 | 9 | 4.0 | 28.0 | 9 | 3.0 | 31.0 | 9 | 4.0 |
| 2B | 312/314/316/318/319 | 0.7 | 17.0 | 9 | 3.0 | 20.0 | 9 | 4.0 | 28.0 | 9 | 3.0 | 31.0 | 9 | 4.0 |
| 2B | 326/332 | 0.5 | 17.0 | 9 | 2.5 | – | – | – | 28.0 | 9 | 2.5 | – | – | – |
| 3B | 302 | 3.0 | 24.0 | 10 | 4.5 | 28.0 | 10 | 5.5 | 35.0 | 10 | 4.5 | 39.0 | 10 | 5.5 |
| 3B | 303/304 | 2.0 | 23.0 | 10 | 4.0 | 27.0 | 10 | 5.5 | 34.0 | 10 | 4.0 | 38.0 | 10 | 5.5 |
| 3B | 305/306/307 | 1.6 | 23.0 | 10 | 3.5 | 27.0 | 10 | 5.5 | 34.0 | 10 | 3.5 | 38.0 | 10 | 5.5 |
| 3B | 308/310 | 1.3 | 22.0 | 10 | 3.5 | 25.0 | 10 | 4.0 | 33.0 | 10 | 3.5 | 36.0 | 10 | 4.0 |
| 3B | 309 | 1.3 | 22.0 | 10 | 3.5 | 25.0 | 10 | 4.0 | 33.0 | 10 | 3.5 | 36.0 | 10 | 4.0 |
| 3B | 309 | 2.0 | 22.0 | 10 | 4.0 | 25.0 | 10 | 5.5 | 33.0 | 10 | 4.0 | 36.0 | 10 | 5.5 |
| 3B | 312/314/316/318 | 0.9 | 21.0 | 10 | 3.0 | 25.0 | 10 | 4.0 | 32.0 | 10 | 3.0 | 36.0 | 10 | 4.0 |
| 3B | 320/322/324/326/330 | 0.7 | 21.0 | 10 | 3.0 | 25.0 | 10 | 4.0 | 32.0 | 10 | 3.0 | 36.0 | 10 | 4.0 |
| 4B | 304 | 3.0 | 33.0 | 12 | 4.5 | 36.0 | 12 | 5.5 | 41.0 | 12 | 4.5 | 45.0 | 12 | 5.5 |
| 4B | 306/307 | 2.0 | 32.0 | 12 | 4.0 | 36.0 | 12 | 5.5 | 41.0 | 12 | 4.0 | 45.0 | 12 | 5.5 |
| 4B | 310 | 1.6 | 32.0 | 12 | 3.5 | 36.0 | 12 | 5.5 | 39.0 | 12 | 3.5 | 43.0 | 12 | 5.5 |
| 4B | 312 | 1.3 | 32.0 | 12 | 3.5 | 36.0 | 12 | 4.0 | 39.0 | 12 | 3.5 | 43.0 | 12 | 4.0 |
| 4B | 316/320/324/330 | 0.9 | 32.0 | 12 | 3.0 | 34.0 | 12 | 4.0 | 39.0 | 12 | 3.0 | 43.0 | 12 | 4.0 |
| 4B | 340/348 | 0.7 | 32.0 | 12 | 3.0 | 34.0 | 12 | 4.0 | 39.0 | 12 | 3.0 | 43.0 | 12 | 4.0 |
| 5B[1] | 302 | 6.0 | 42.0 | 18 | 7.5 | – | – | – | 70.0 | 18 | 7.5 | – | – | – |
| 5B[1] | 304 | 4.0 | 47.0 | 18 | 5.5 | 50.0 | 18 | 7.0 | 75.0 | 18 | 5.5 | 78.0 | 18 | 7.0 |
| 5B[1] | 310 | 3.0 | 47.0 | 18 | 4.5 | 50.0 | 18 | 7.0 | 75.0 | 18 | 4.5 | 78.0 | 18 | 7.0 |
| 5B[1] | 314/316 | 2.0 | 46.0 | 18 | 4.0 | 49.0 | 18 | 5.5 | 74.0 | 18 | 4.0 | 77.0 | 18 | 5.5 |
| 5B[1] | 320 | 1.6 | 46.0 | 18 | 3.5 | 49.0 | 18 | 5.5 | 74.0 | 18 | 3.5 | 77.0 | 18 | 5.5 |
| 5B[1] | 330/340/348 | 1.3 | 45.0 | 18 | 3.5 | 48.0 | 18 | 4.0 | 74.0 | 18 | 3.5 | 77.0 | 18 | 4.0 |
| 5B[1] | 350/354/364 | 0.9 | 45.0 | 18 | 3.0 | 48.0 | 18 | 4.0 | 74.0 | 18 | 3.0 | 77.0 | 18 | 4.0 |

- M4  straight plug, short version, clamping type D or M (solder or crimp contacts)

| Series | Type | ø contact A (mm) | M4 Solder L | M4 Solder S | M4 Solder T | M4 Crimp L | M4 Crimp S | M4 Crimp T |
|---|---|---|---|---|---|---|---|---|
| 0B | 302/303 | 0.9 | 9.5 | 8 | 3.0 | 13.0 | 8 | 4.0 |
| 0B | 304/305 | 0.7 | 9.5 | 8 | 3.0 | 13.0 | 8 | 4.0 |
| 0B | 306/307/309[2] | 0.5 | 10.0 | 8 | 2.5 | 13.5 | 8 | 3.0 |

L: ± ... mm; S: ± 0.5 mm; T: ± 0.2 mm.
... increased by 2 mm for the largest collet (D56 in 0B series; D76 in 1B series).
... ised by 13 mm for the largest collet (D25).
... with male contacts.

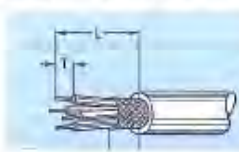

Headquarters
European XFEL GmbH
Albert-Einstein-Ring 19
22761 Hamburg
Germany/
www.xfel.eu

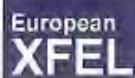

Kai-Erik Ballak
WP-75 (Detector Development)
Electronics technician

Phone +49 40 8998-6430
Fax +49 40 8998-1905
kai-erik.ballak@xfel.eu

Mailing address
European XFEL GmbH
Notkestraße 85
22607 Hamburg
Germany

## C.5 Specification of Translation Motor

### 4.094 Precision Linear Stage PLS-85 Vacuum

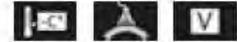

**FACTS**

| Load characteristics | Fx(N) | Fy(N) | Fz(N) | Mx(Nm) | My(Nm) | Mz(Nm) | kax(μrad/Nm) | kay(μrad/Nm) |
|---|---|---|---|---|---|---|---|---|
| FV | 60 | 50 | 100 | 25 | 30 | 20 | 70 | 40 |
| HV | 60 | 50 | 100 | 25 | 30 | 20 | 70 | 40 |
| UHVG | 60 | 50 | 100 | 25 | 30 | 20 | 70 | 40 |

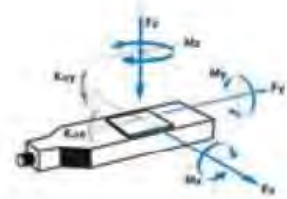

especially suited for high precision applications with limited space conditions while allowing loads of up to 10 kg. Cross-roller bearings guarantee very high guiding stiffness. Driven by a re-circulating ball screw with 1 mm pitch. The PLS-85 Vacuum can be mounted in any orientation. For demanding positioning tasks, the PLS-85 Vacuum linear stages can be supplied with a side-mounted linear scale. The PLS-85 Vacuum is equipped with a 2-phase steppervacuum motor and has two limit switches.

Please see "GLOSSARY Vacuum Specification" for all vacuum specification.

The PLS-85 Vacuum linear stages are

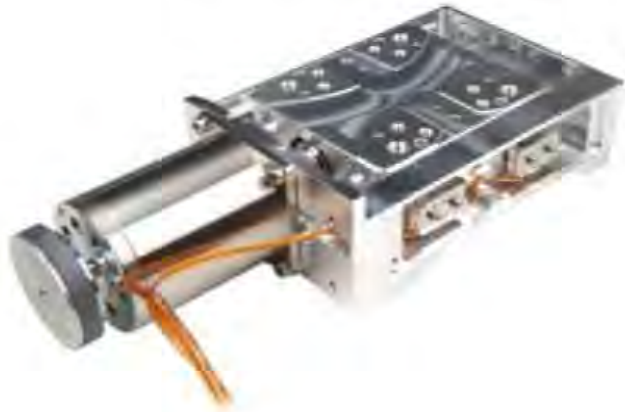

**KEY FEATURES**

- Vacuum up to 10-9 hPa
- Travel range up to 155 mm (6")
- Uni-directional repeatability down to 0.05 μm
- Maximum speed 6 mm/sec
- Load capacity up to 8 kg
- Integrated switches
- Option: linear scale

**PI|miCos**

**TECHNICAL DATA**

| | | | | | |
|---|---|---|---|---|---|
| Travel range (mm) | 26 | 52 | 102 | 155 | |
| Straightness / Flatness (μm) | ±1 | ±2 | ±4 | ±6 | |
| Pitch (μrad) | ±60 | ±90 | ±120 | ±150 | |
| Yaw (μrad) | ±60 | ±80 | ±100 | ±130 | |
| Weight (kg) | 0.9 | 1.2 | 1.5 | 1.8 | |
| Vacuum type | FV | HV | UHVG | | |
| Linear scale | | | | FV | UHV |
| Speed max. (mm/sec) | 6 | 2.5 | 2.5 | | |
| Resolution calculated (μm) | 5 (FS) | 5 (FS) | 5 (FS) | 0.05 | 0.05 |
| Resolution typical (μm) | 0.05 | 0.05 | 0.05 | 0.05 | 0.05 |
| Bi-directional Repeatability (μm) | ±1 | ±1 | ±1 | ±0.1 | ±1 |
| Uni-directional Repeatability (μm) | 0.1 | 0.1 | 0.1 | 0.05 | 0.05 |
| Nominal Current (A) | 1.2 | 1.2 | 1.2 | | |

| | |
|---|---|
| Accuracy | on request |
| Velocity range (mm/sec) | 0.001 ... 6 |
| Material | Aluminum or Stainless steel, suitable for vacuum |

Vacuum Note: FS = full step, RE = rotary encoder
More info: Detailed information concerning motors and encoders, see appendix.

## C.5 Specification of Rotation Motor

**5.132 Rotation Stage RS-40 Vacuum** 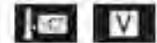

### FACTS

| Load characteristics | FX(N) | FY(N) | MX(Nm) | MZ(Nm) | kαX(μrad/Nm) |
|---|---|---|---|---|---|
| ~~FV~~ | ~~5~~ | ~~10~~ | ~~1~~ | ~~02~~ | ~~270~~ |
| 10-6 | 5 | 10 | 1 | 02 | 270 |
| HV | 5 | 10 | 1 | 02 | 270 |
| UHVG | 5 | 10 | 1 | 02 | 270 |

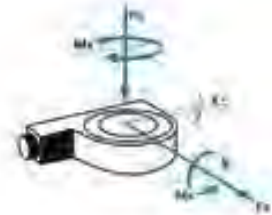

The RS-40 Vacuum rotation stage is very compact but offers a big 20 mm (25 mm holding diameter) aperture. A precision bearing guarantees a perfectly smooth move. The RS-40 Vacuum rotation stages have nearly zero backlash worm gear reduction. All RS-40 Vacuum motorized rotation stages are equipped with a reference switch and are offered with a special 2-phase geared vacuum stepper motor.

Please see 'GLOSSARY Vacuum Specification' for all vacuum specification.

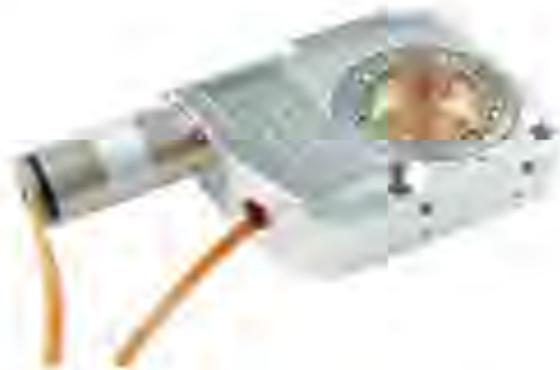

### KEY FEATURES

- Vacuum up to $10^{-9}$ hPa
- Clear aperture 20 mm
- Uni-directional repeatability down to 0.005°
- Maximum speed 1.5°/sec
- Load capacity up to 1 kg
- Integrated hall reference switch
- Optionally rotary encoder on the rotation axis

### TECHNICAL DATA

| Travel range (°) | 360 | | | |
|---|---|---|---|---|
| Flatness (Bearings) (μm) | ±5 | | | |
| Eccentricity (Bearings) (μm) | ±5 | | | |
| Wobble (Bearings) (μrad) | ±35 | | | |
| Weight (kg) | 0.4 | | | |
| Vacuum type | FV | 10-6 | HV | UHVG |
| Speed max. (°/sec) | 1.5 | 0.6 | 0.6 | 0.27 |
| Resolution calculated (°) | 0.00219Δ (FS) | 0.00219Δ (FS) | 0.02 (FS) | 0.02 (FS) |
| Resolution typical (°) | 0.005 | 0.005 | 0.002 | 0.002 |
| Bi-directional Repeatability (°) | ± 0.04 | ± 0.04 | ± 0.03 | ± 0.03 |
| Uni-directional Repeatability (°) | 0.005 | 0.005 | 0.005 | 0.005 |
| Nominal Current (A) | 0.25 | 0.25 | 1.2 | 1.2 |
| Accuracy | on request | | | |
| Velocity range (mm/sec) | 0.002...1.5 | | | |
| Material | Aluminum, stainless steel, red brass | | | |

Vacuum Note: FS = Full step, RE = rotary encoder
More info: Detailed information concerning motors and encoders see appendix.

**PI|miCos**